\documentclass[%
 reprint,
 amsmath,amssymb,
 aps,
]{revtex4-2}

\usepackage{graphicx}% Include figure files
\usepackage{dcolumn}% Align table columns on decimal point
\usepackage{bm}% bold math
\usepackage{color}
\usepackage{pgfplots,mathtools}
\usepackage{hyperref}

\usepackage{braket}
\usepackage{slashed}
\usepackage[compat=1.0.0]{tikz-feynman}
\newcommand{\be}{\begin{equation}}
\newcommand{\ee}{\end{equation}}
\newcommand{\bea}{\begin{eqnarray}}
\newcommand{\eea}{\end{eqnarray}}
\newcommand{\ba}[1]{\begin{array}{#1}}
\newcommand{\ea}{\end{array}}
\newcommand{\nn}{\nonumber}

\newcommand{\ep}{\epsilon}

\begin{document}
    \title{Wiedemann-Franz law violation domain for graphene and nonrelativistic systems}
    \author{Thandar Zaw Win, Cho Win Aung, Gaurav Khandal  \\
    Sabyasachi Ghosh}
%and Sesha Pawan Vempati}
    \affiliation{Department of Physics, Indian Institute of Technology Bhilai, Kutelabhata, Durg 491002, India}
%
%\maketitle
    \begin{abstract}
   {A systematic non-fluid to fluid transition framework and comparative research on Lorenz ratios for graphene and nonrelativistic systems have been studied to identify their Wiedemann-Franz law violation domain. Here, Lorenz ratio is defined as thermal conductivity divided by electrical conductivity times temperature times Lorenz number. In non-fluid framework, Lorenz ratio become exactly one, which means that the Wiedemann-Franz is obeyed within a Fermi Liquid domain. When one enters from Fermi Liquid to Dirac Fluid domain, Lorenz ratio becomes less than one in non-fluid framework but in fluid framework, it always remain greater than one for both domain. By compiling our outcomes and connecting with experimental data, a non-fluid to fluid transition framework is expected during the transition from Fermi Liquid to Dirac Fluid domain.}
    \end{abstract}
\maketitle
    \section{Introduction}
    %Graphene \cite{katsnelson2007graphene} is a thin, two dimensional layer of carbon atoms arranged in a hexagonal lattice, there are many exciting things such as zero band gap  of the Dirac fluid region \cite{crossno2016observation}, the energy dispersion relation \cite{katsnelson2007graphene}, and the Quantum-Hall effect \cite{novoselov2005two}.
%
    In 1853, Gustav Wiedemann and Rudolph Franz experimentally discovered a universal or constant value of the ratio between thermal ($\kappa$) and electrical ($\sigma$) conductivity, which are approximately followed by all metals. Later, in 1872, Ludvig Lorenz theoretically realized that the ratio in terms of the universal constants $k_B$ (Boltzmann constant) and $e$ (electric charge) is:
    \bea 
    \frac{\kappa}{\sigma T} &=& \frac{\pi^2}{3}\left(\frac{k_B}{e}\right)^2
    \nn\\
    &=& L_0 = 2.445 \times 10^{-8} watt\frac{\Omega}{K^2}~,
    \eea 
    where $L_0$ is known as the Lorenz number. The fact is very well known from our text book Refs.~\cite{ashcroft2022solid,devanathan2021wiedemann,Pathria:1996hda,pillai2006solid,puri1997solid}. In natural unit, $\hbar=c=k_B=1$ and $e^2=\frac{4\pi}{137}$, we can express temperature $T$ in eV, $\sigma$ in eV and $\kappa$ in eV$^2$. So Lorenz number will be a dimensionless ratio $L_0=\frac{137\pi}{12}\approx 35.84$. In the present paper, we will mostly use the natural unit methodology for convenience, but we may go to the actual unit in some cases (whenever necessary).  
%
    %In 1897, J.J.Thomson discovered the electron. After the discovery of electrons, Drude proposed the concept of electrical conductivity in metals in 1900, and he proposed that the conduction in metals is because of the flow of electrons. In 1909, Lorenz extended this theory, known as the Drude-Lorenz theory, and In 1928, Sommerfeld applied the quantum theory on electrons. References are
    %\cite{ashcroft2022solid}, \cite{devanathan2021wiedemann}, \cite{Pathria:1996hda}, \cite{pillai2006solid}, \cite{puri1997solid}, \cite{novoselov2004electric}.\\

    The Wiedemann-Franz (WF) law has proven remarkably robust for many metallic systems, where electrons are the main electric and thermal charge transportation carriers. Due to the similar mechanism of transportation in free electron theory~\cite{ashcroft2022solid,devanathan2021wiedemann,Pathria:1996hda,pillai2006solid,puri1997solid}, the dimensionless (in natural unit) ratio of two transport coefficients - thermal and electrical conductivity becomes constant. However, deviations have been observed in many systems like $Li_{0.9}Mo_6O_{17}$ \cite{wakeham2011gross}, $CeCoIn_5$ \cite{tanatar2007anisotropic}, $ZrZn_2$ \cite{smith2008marginal}, $YbRh_2Si_2$ \cite{pfau2012thermal}, $(Pr,Ce)_2CuO_4$ \cite{hill2001breakdown} and $VO_2$ \cite{lee2017anomalously}. For our convenience, if we define the ratio of thermal and electrical conductivity as $L=\frac{\kappa}{\sigma T}$, and if we call $L/L_0$ as Lorenz ratio, then the validity of the WF law will be established from the relations $L=L_0$ or $\frac{L}{L_0}=1$.  On the other hand, relations $\frac{L}{L_0}>1$ or $\frac{L}{L_0}<1$ reflect violation of the WF law. A large  diverging outcome  $\frac{L}{L_0} > 10^4$ is expected in one-dimensional exotic Tomonaga-Luttinger liquids~\cite{wakeham2011gross}. On the other hand, a strong downward violation of the WF Law $L<L_0$ was observed in $WP_2$ semimetal~\cite{WP2_1_departure,WP2_2_departure} and MoP binary compound~\cite{MoP_departure}. This downward violation is also observed in heavy fermion metals~\cite{tanatar2007anisotropic}, marginal ferromagnetic metal~\cite{smith2008marginal}, anti-ferromagnetic metal~\cite{pfau2012thermal} and copper-oxide superconducting material~\cite{hill2001breakdown} at low-temperature regions as well as in metallic vanadium dioxide~\cite{lee2017anomalously} at high-temperature range ($240$-$340^\circ$K). Similar violation is also observed for the nuclear matter, produced in heavy ion collision experiments~\cite{VDWHRG_2023, CSPM_2019, HRG1_2019, HRG2_2023}. Understanding the WF law violation mechanism of that wide range of systems is a non-trivial task and challenge for the theoretical community to explain. 
    %\textcolor{blue}{The violation of the law in ultraclean conductors \cite{PhysRevLett.115.056603} by considering the relaxation time for electron liquids and the violation for 2D and 3D Fermi liquids \cite{PhysRevB.99.085104} by considering phonon and impurity scatterings have also observed in some theoretical calculations}. 

    Recently, Crossno \textit{et al.}~\cite{crossno2016observation} have found a similar kind of the WF law violation in graphene systems by tuning the doping concentration and temperature. Unlike standard metals, the Fermi level of graphene can be moved upwards or downwards from the relative Dirac point depending upon the n- or p-type doping of graphene. Fermi energy $(\epsilon_F)$ or a chemical potential $(\mu)$ at Dirac point is considered zero for undoped graphene and, via n- or p-type doping, will shift it towards the positive or negative direction. The net charge carrier concentration will be tuned from zero to non-zero when one experimentally goes from undoped to doped graphene cases, and theoretically, one goes from $\epsilon_F=0$ to $\epsilon_F\neq 0$ cases. This kind of tuning possibility of $\epsilon_F$ can not be expected for metal systems as it almost remains constant. Their typical values remain within the range  $\epsilon_F=2-10$~eV, for which one can expect the limit $\epsilon_F/T>>1$ at room temperature $T\approx 0.025$ eV. In this limit, one can consider electrons in metal as Fermi gas and use Sommerfeld expansion, which provides a linear T dependent on electron-specific heat. Fermi gas is completely based on the crude non-interacting assumption, but there is a theory of interacting Fermions system, which is popularly known as the Fermi Liquid (FL) Theory\cite{gochan2020fermi}. Originally, Landau~\cite{Landau1, Landau2, Landau3} proposed this phenomenological theory for studying $^3$He. In FL prescription, interaction is taken care of via the effective mass of electrons by assuming a quasi-particle picture.  So, if we define $\epsilon_F/T>1$ as the FL domain and $\epsilon_F/T>>1$ as the Fermi gas (FG) domain, then electron-doped graphene systems almost follow WF law. When we go towards undoped or clean graphene systems with $\epsilon_F/T<1$, Fermi liquid theory becomes invalid (as it is commonly accepted). It is concluded from the experimental observation of WF law violation in this $\epsilon_F/T<1$ or $\epsilon_F/T<<1$ domains, popularly called Dirac fluid (DF) or Dirac Liquid (DL) domain. Theoretical works~\cite{rycerz2021wiedemann,tu_Yin_23_the,Theory_G_WF,Mahajan:2013cja,Lucas:2018kwo,zarenia2019breakdown, PhysRevLett.115.056603,PhysRevB.99.085104} are attempted to explain this WF law violation in the DF domain. In this regards, electron hydrodynamics (eHD) or fluid dynamics in graphene system, recently observed by Refs~\cite{exp1,exp2, PRB_21, PRB_2_21, PRB_3_21, PRR, Nat_Com_21, PRB_coll, Nat_19, Sci_19, Hall_Sci_19, Nat_Tec_19, Nat_Com_18, Sci_16_Res}, is thought to be linked with this WF law violation. In the direction of material perspective \cite{principi2015violation}, one can link eHD and WF law violation as follows. The electron-electron scattering length becomes larger than the electron-impurity or electron-phonon scatterings in DF domain but reverse is occured in FL or Fermi gas domain. We know that the relaxation time is inverse to scattering strength. So, the e-e scattering dominating domain i.e. DF domain shows eHD behaviour, where electron relaxation time becomes much much smaller than the system size. In this context, the present work is planned to explore a transition from non-fluid to fluid type expressions of Lorenz ratio for a 2-dimensional (2D) graphene (G) system with linear energy-momentum relation ($E\propto p$) as well as a 3-dimensional (3D) metal system with nonrelativistic (NR) energy-momentum relation ($E\propto p^2$). For mathematical completeness, we have demonstrated all possible cases like 2D-NR, 3D-NR, 2D-G, and 3D-G but our final destination is kept to show how fluid expressions of 2D-G case can be associated with WF law violation and how non-fluid expressions of 3D-NR case, followed by metal systems, obey WF law. In the direction of eHD, our earlier comparative research on ideal~\cite{TZW} and viscous dissipating~\cite{CWA} parts of energy-momentum tensor calculation clearly demonstrate that the graphene hydrodynamics is neither relativistic hydrodynamics nor nonrelativistic hydrodynamics. The present work has gone through similar comparative research on electrical conductivity, thermal conductivity, and Lorenz ratio for a systematic search of the WF law violation domain. 

    The article is organized as follows, in Sec. \ref{sec:sec2}, the formalism part of WF law calculations for different cases. The results and brief interpretation are in Sec. \ref{sec:sec3} and Sec. \ref{sec:sec4} respectively. At the end, our study is concluded in Sec. \ref{sec:sec5}.

%%%%%%%%%%%%%%%%
    \section{Formalism }
    \label{sec:sec2}
    Graphene is a 2-dimensional single atomic layer of carbon atoms tightly bound into a honeycomb lattice~\cite{geim2007rise}. Near the Dirac point, electrons in graphene follow the linear dispersion relation
    \begin{equation}
     \epsilon\left(k\right) = pv_F,
    \label{nanu}
    \end{equation}
    where $v_F$ is the Fermi velocity of electrons in graphene, whose values remain within the range $0.003c$-$0.01c$~\cite{vg_nature}. 
    In one direction, electrons in graphene do not follow a quadratic relation between energy and momentum ($\epsilon = p^2/(2m)$) like in the nonrelativistic case. On the other hand, its velocity ($v_F$) is not close to the speed of light ($c$), so no relativistic corrections are expected. In that sense, electrons in graphene follow neither relativistic nor nonrelativistic. So, its different properties may not be the same as expected for traditional nonrelativistic or relativistic matters as shown in Refs.~\cite{TZW,CWA}, for thermodynamics~\cite{TZW} and dissipative viscous~\cite{CWA} properties. In this regard, thermal and electrical conductivity for the graphene (G) case may also be interesting to compare with nonrelativistic (NR) and ultrarelativistic (UR) cases. 

    \subsection{Non-fluid description}
    \label{sec:NF}
    For systematic study and understanding, we will calculate the Lorenz ratios for the 3D system and the 2D system. The expressions for 2D graphene are derived and analyzed in this section, highlighting its distinct transport characteristics. Meanwhile, the corresponding calculations for other systems, including 3D non-relativistic (NR), 3D graphene (G), and ultra-relativistic (UR) cases in both 3D and 2D, are provided in Appendix \ref{Appendix_D} for completeness. Then, we will also examine the limiting conditions using Sommerfeld expansion.

   {\bf 2D-G:} For actual 2D graphene case, the electrical and thermal transport properties are:
    \begin{equation}
   	\sigma_{G}^{2D} = \frac{ne^2 \tau_c v_g^2}{\mu}~,
   	\label{Sig_NF}
\end{equation}
 \begin{equation}
	\sigma_{Q}^{3D} = \frac{ne^2 \tau_c c^2}{\mu_q}~,
	\label{Sig_NF}
\end{equation}

\begin{equation}
	\kappa_{G}^{2D} = \frac{1}{2}nv_F^2 \tau_c \prescript{2D}{G}{[C_V]_e}~,
	\label{kapa_NF}
\end{equation}

 the ratio of $\kappa_{G}^{2D}$ and $\sigma_G^{2D}$ is given by
    \begin{equation}
    \frac{\kappa_{G}^{2D}}{\sigma_{G}^{2D}} = \frac{1}{2}\frac{\epsilon_F}{e^2} \prescript{2D}{G}{[C_V]_e}.
    \label{body3}
    \end{equation}
\begin{equation}
	\frac{\kappa_{G}^{2D}}{\sigma_{G}^{2D}} = \frac{1}{2}\frac{\epsilon_F}{e^2} \prescript{2D}{G}{[C_V]}.
	\label{body3}
\end{equation}
 the ratio of $\kappa_{Q}^{3D}$ and $\sigma_Q^{3D}$ is given by
\begin{equation}
	\frac{\kappa_{Q}^{3D}}{\sigma_{Q}^{3D}} = \frac{1}{3}\frac{\mu_q}{e^2} \prescript{3D}{Q}{[C_V]_e}.
	\label{body3}
\end{equation}
\begin{equation}
	\frac{\kappa_{Q}^{3D}}{\sigma_{Q}^{3D}} = \frac{1}{3}\frac{\mu_q}{e^2} \prescript{3D}{Q}{[C_V]}.
	\label{body3}
\end{equation}
    Using two possible expressions (see Appendix \ref{Appendix_C}) of specific heat for 2D-G system,
    \begin{equation}
    \prescript{2D}{G1}{[C_{V}]_{e}} = 2 k_B\Bigg[3 \frac{f_3\left(A\right)}{f_2\left(A\right)} - 2 \frac{f_2\left(A\right)}{f_1\left(A\right)} \Bigg], 
    \label{1jay3}
    \end{equation}
 
and

 \begin{equation}
	\prescript{2D}{G2}{[C_{V}]_{e}} = 2 k_B\Bigg[3 \frac{f_3\left(A\right)}{f_2\left(A\right)} - \frac{\ep_F}{k_BT} \Bigg], 
	\label{1sawanra3}
\end{equation}

    in Eq.~(\ref{body3}), we get the Lorenz ratios
    \begin{equation}
    \frac{L_{G1}^{2D}}{L_0} =  \frac{\kappa_{G}^{2D}}{\sigma_{G}^{2D} T L_0} = \frac{3}{\pi^2} \frac{\epsilon_F}{k_BT} \Bigg[3 \frac{f_3\left(A\right)}{f_2\left(A\right)} - 2 \frac{f_2\left(A\right)}{f_1\left(A\right)} \Bigg] 
    \label{jay5}
    \end{equation}

    and
    \begin{equation}
    \frac{L_{G2}^{2D}}{L_0} =  \frac{\kappa_{G2}^{2D}}{\sigma_{G2}^{2D} T L_0} = \frac{3}{\pi^2} \frac{\epsilon_F}{k_BT} \Bigg[ 3\frac{f_3\left(A\right)}{f_2\left(A\right)} - \frac{\epsilon_F}{k_B T}\Bigg]  
    \label{jay6}
    \end{equation}
    respectively. The SL of Eqs.~(\ref{jay5}) and (\ref{jay6}) will be
    \begin{equation}
    L_{G1}^{2D} = \frac{\pi^2}{3}\left(\frac{k_B}{e}\right)^2 = L_0,
    \label{jay5SL}
    \end{equation}
    and
    \begin{equation}
    L_{G2}^{2D} = \frac{2\pi^2}{3}\left(\frac{k_B}{e}\right)^2 = 2 L_0
    \label{jay6SL}
    \end{equation}
    respectively.

 %%%%%%%%%%%%%%%%%%%%%%%%%%%%%%%%%%%%%%%%%%%%%%%%%%%

    \subsection{An approach towards the fluid description}
    \label{sec:F}
    The previous subsection provides the expression of $\kappa$ and $\sigma$ using a standard solid state physics framework, where no fluid concept has been entered. So, we can call that the non-fluid description. In the present section, we will try to build an approach towards a fluid description by using relaxation time approximation-based Boltzmann's transport equations. Here, we are using the terminology "towards the fluid description" because the actual phenomenological fluid description \cite{Theory_G_WF} for explaining the Wiedemann-Franz law violation in the experimental 2D graphene system is quite rigorous. Aim of the present work is to zoom in the intermediate journey from non-fluid to fluid description. Interestingly, we will get a Wiedemann-Franz law violating domain in $T-\ep_F$ plane for all cases. Let us first address the brief calculations of the Lorenz ratio for 2D-G systems; then, we will write down its final expressions for other possible systems like 2D-NR, 3D-NR, and 3D-G instead of repeating similar calculations  (see Appendix \ref{Appendix_E}).
    %\subsubsection{The Electrical Conductivity}
    %Let us have a sample of graphene monolayer in the $xy-$ plane and an electric field E in $x-$direction applied to it. In the applied electric field, the electrons feel a force and move in the opposite direction of the field with a constant drift velocity after just a minimal time. Due to this external force, there is dissipation in the equilibrium system, and the system becomes non-equilibrium, and after some time, it goes into a steady state. After just removing the field, the system again comes into an equilibrium state, and in that time, it comes into an equilibrium state from the steady state is known as relaxation time($\tau_c$). So in the steady state, by using 

    {\bf 2D-G:} Let us assume a local thermalization picture of electron fluid in a 2D-G system, where the equilibrium distribution function (see Appendix \ref{Appendix_A})
    \begin{equation}    
    f_0\left(T(x^\mu) \right) = \frac{1}{e^{\left(\epsilon - \epsilon_F\right)/k_BT(x^\mu)} + 1},
    \label{eq:f0_T}
    \end{equation}

will be $x^\mu$-dependent due to temperature profile $T(x^\mu)$. Here, $x^\mu=(ct, x^i)$, a four-dimensional coordinate is considered for general notation, but we have to take care of $i=1,2$ for the 2D system and $i=1,2,3$ for the 3D system.

    Now, let us first write down the macroscopic definitions for electrical and thermal conductivity
    \bea
    {\vec J}&=&\sigma {\vec E},
    \nn\\
    {\vec Q}&=&\kappa {\vec \nabla}T~,
    \label{eq:J_E}
    \eea
    where electrical current vector ${\vec J}$ and heat flow vector ${\vec Q}$ can have microscopic expressions:
    \bea
    {J_i}&=&ge\int\frac{d^2p}{h^2} v_i \delta f_\sigma,
    \nn\\
    {Q_i}&=&g\int\frac{d^2p}{h^2} \epsilon v_i \delta f_\kappa.
    \label{eq:J_df}
    \eea
    Here, we are assuming that the external electric field $E_i$ and the temperature gradient ${\vec \nabla}T$ will create deviations $\delta f_\sigma$ and $\delta f_\kappa$ respectively from the equilibrium distribution $f_0$.

    It is relaxation time approximation (RTA) based on Boltzmann's transport equation (BTE)~\cite{BTE}
    \begin{equation}
    \frac{\partial f}{\partial t}+ \frac{\partial x^i}{\partial t} \frac{\partial f}{\partial x^i}+ \frac{\partial p^i}{\partial t} \frac{\partial f}{\partial p^i}=-\frac{ \delta f}{\tau_c}~,
    \label{eq:BTE}
    \end{equation}
    which will guide us to guess appropriate form of $\delta f_\sigma$ and $\delta f_\kappa$.
    Considering $f=f_0+\delta f\approx f_0$ in left hand side of Eq.~(\ref{eq:BTE}) and local thermalization assumption of $f_0$, given in Eq.~(\ref{eq:f0_T}), we can simplify Eq.~(\ref{eq:BTE}) as
    \bea
    v^i \frac{\partial f_0}{\partial x^i}+ eE^i \frac{\partial f_0}{\partial p^i}&=&-\frac{ {\delta f}_\sigma}{\tau_c}-\frac{ {\delta f}_\kappa}{\tau_c}
    \nn\\
    v^i \frac{\partial T}{\partial x^i}\frac{\partial\epsilon}{\partial T}\frac{\partial f_0}{\partial \epsilon}+ eE^i \frac{\partial \epsilon}{\partial p^i}\frac{\partial f_0}{\partial \epsilon}&=&-\frac{ {\delta f}_\sigma}{\tau_c}-\frac{ {\delta f}_\kappa}{\tau_c}
    \nn\\
    v^i \frac{\partial T}{\partial x^i}[C_V]_e\frac{\partial f_0}{\partial \epsilon}+ eE^i v_i\frac{\partial f_0}{\partial \epsilon}&=&-\frac{ {\delta f}_\sigma}{\tau_c}-\frac{ {\delta f}_\kappa}{\tau_c}.
    \label{eq:BTE2}
    \eea
    Here, we consider the approximation $[C_V]_e\approx \frac{\partial\epsilon}{\partial T}$, and one can expect again two possible definitions of specific heat as discussed in the earlier section.
    From Eq.~(\ref{eq:BTE2}), we can get the form of $\delta f_\sigma$ and $\delta f_\kappa$ as
    \bea 
    \delta f_\sigma &=& eE^i v_i \left(- \frac{\partial f_0}{\partial \epsilon}\right)\tau_c,
    \nn\\
    \delta f_\kappa &=& v^i \left(-\frac{\partial f_0}{\partial \epsilon}\right) [C_V]_e \left(\frac{\partial T}{\partial x^i}\right) \tau_c,
    \label{eq:df}
    \eea 
    \begin{equation}
    	\delta f_\kappa = v^i \left(-\frac{\partial f_0}{\partial \epsilon}\right) [C_V] \left(\frac{\partial T}{\partial x^i}\right) \tau_c
\end{equation}
    with
    \be 
    \left(- \frac{\partial f_0}{\partial \epsilon}\right)=\beta f_0(1-f_0).
    \ee 
    Using Eq.~(\ref{eq:df}) in Eq.~(\ref{eq:J_df}) and then comparing with Eq.~(\ref{eq:J_E}), we will get the final expressions of electrical and thermal conductivity:
    \bea
    \sigma &=&  g e^2  \frac{v_F^2}{2} \tau_c\int \frac{d^2p}{h^2} \beta f_0(1-f_0)
    \nn\\
    \implies\sigma_G^{2D} &=& 2\pi k_B \tau_c \left(\frac{e}{h}\right)^2 f_1\left(A\right) T,
    \label{sig_towards}
    \eea
    and
    \bea
    \kappa &=& g \epsilon  \frac{v_F^2}{2} \tau_c [C_V]_e\int \frac{d^2p}{h^2} \beta f_0(1-f_0)
    \nn\\
    \implies\kappa_G^{2D} &=& \frac{4 \pi k_B^2\tau_c }{h^2} [C_V]_{e}\,f_2\left(A\right) T^2.
    \label{manohar}
    \eea
    \begin{equation}
    	\kappa_{F/NF}^{G}=\frac{\tau_c }{\pi} [C_V]_{F/NF}\,f_2\left(A\right) T^2.
    \end{equation}
    \begin{equation}
    	\kappa_{F/NF}^{Q}=\frac{24 \ \tau_c }{\pi^2} [C_V]_{F/NF}\,f_3\left(A\right) T^3.
    \end{equation}
    Now, using two different forms of specific heat, given in Eqs.~(\ref{1jay3}) and (\ref{1sawanra3}) in Eq.~(\ref{manohar}),
    we get
    \be 
    \kappa_{G1}^{2D} = \frac{8\pi k_B^3 \tau_c }{h^2} \Bigg[ 3\frac{f_3\left(A\right)}{f_2\left(A\right)} - 2 \frac{f_2\left(A\right)}{f_1\left(A\right)} \Bigg] f_2\left(A\right) T^2,
    \label{K_NF_CV_1}
    \ee 
    and
    \be 
    \kappa_{G2}^{2D} = \frac{8\pi k_B^3 \tau_c}{h^2} \Bigg[ 3\frac{f_3\left(A\right)}{f_2\left(A\right)} - \frac{\epsilon_F}{k_B T} \Bigg] f_2\left(A\right) T^2,
    \ee 
    and hence, corresponding Lorenz ratios will be
    \be 
    \frac{L_{G1}^{2D}}{L_0} = \frac{12}{\pi^2} \Bigg[3 \frac{f_3\left(A\right)}{f_2\left(A\right)} - 2 \frac{f_2\left(A\right)}{f_1\left(A\right)} \Bigg] \frac{f_2\left(A\right)}{f_1\left(A\right)},
    \label{L2DG1}
    \ee 
    and
    \be 
    \frac{L_{G2}^{2D}}{L_0} = \frac{12}{\pi^2} \Bigg[ 3\frac{f_3\left(A\right)}{f_2\left(A\right)} - \frac{\epsilon_F}{k_BT}\Bigg]\frac{f_2\left(A\right)}{f_1\left(A\right)}
    \label{L2DG2}
    \ee 
    respectively.

    \section{Results}
    \label{sec:sec3}
    \begin{figure*}
    \centering
    \includegraphics[scale= 0.3]{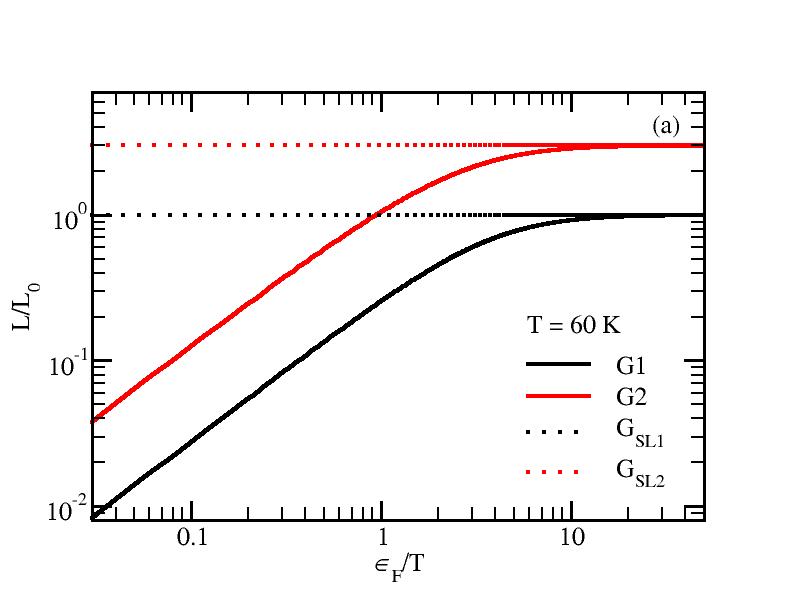}
    \includegraphics[scale= 0.3]{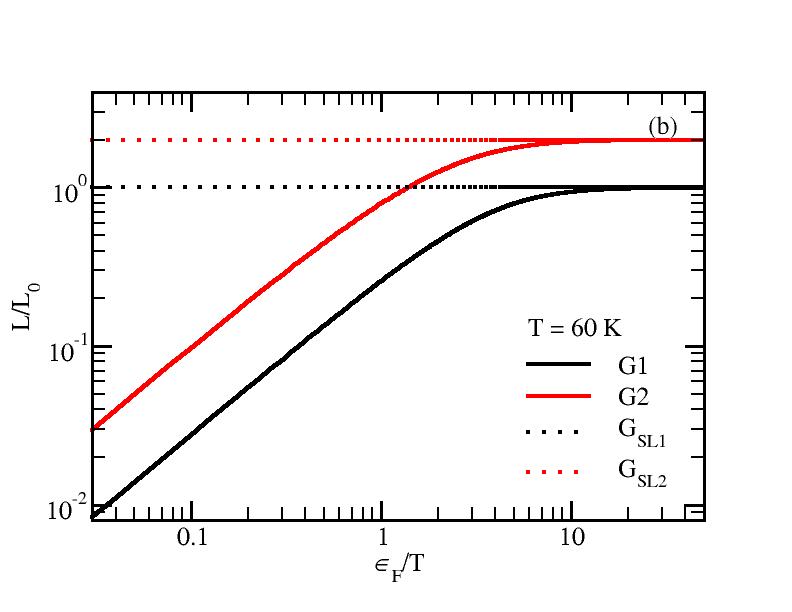}
    \caption{The Lorenz ratio regarding the chemical potential for non-fluid descriptions of graphene case (a) for 3D and (b) for 2D expressions}
    \label{fig:NF_G}
    \end{figure*}
%    \vspace{4.25cm}
    \begin{figure*}
    \centering
    \includegraphics[scale= 0.3]{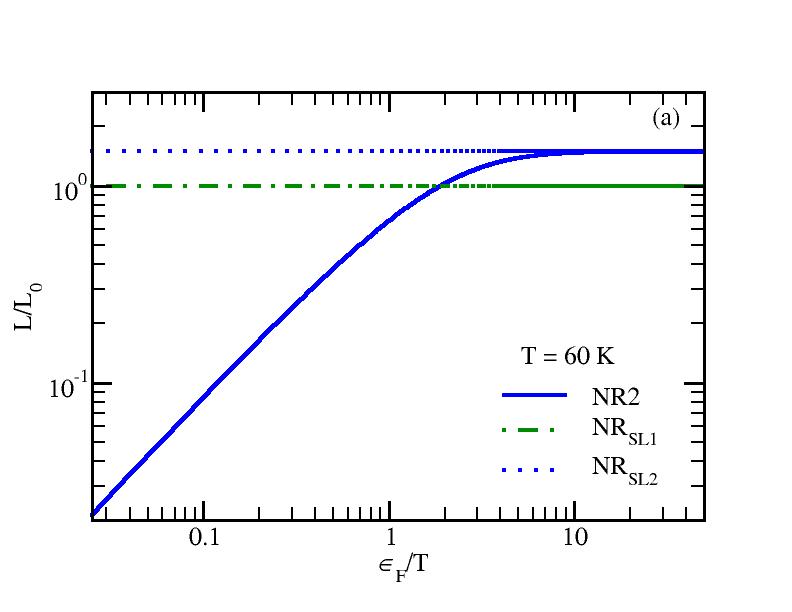}
    \includegraphics[scale= 0.3]{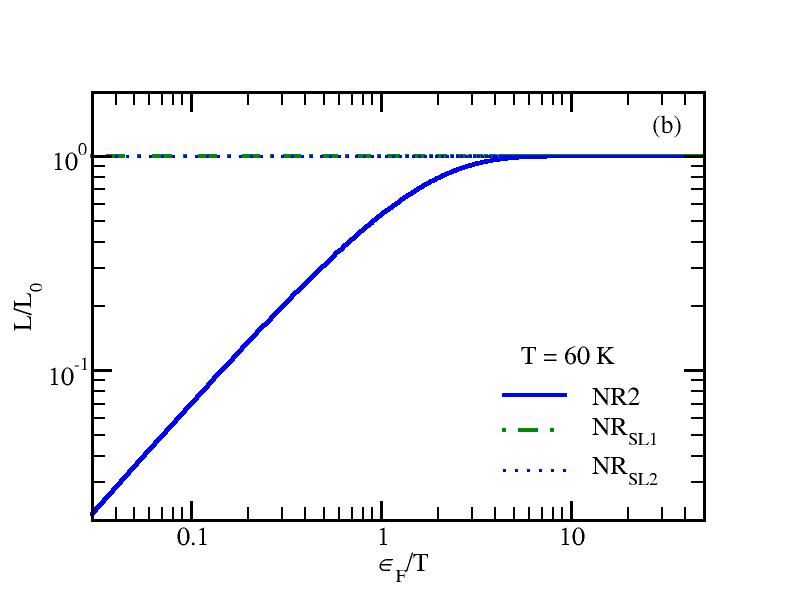}
    \caption{The Lorenz ratio regarding the chemical potential for non-fluid descriptions of nonrelativistic case (a) for 3D and (b) for 2D expressions}
    \label{fig:NF_NR}
    \end{figure*}
    In the above section, we have calculated all the general expressions of electrical and thermal conductivity and their ratios to check the validity of the Wiedemann-Franz law fore 2D graphene (G) system. Using those final expressions, the present section is intended to go with their numerical estimations. During the result generation, we used the natural unit $k_B=\hbar=c=1$ for our convenience. Dimensionless quantities Lorenz ratio ($L/L_0$) and $\epsilon_F/T$ will be taken as the y and x axes of all graphs.

%%%%%%%%%%%%%%%%%%%(L/L0_Electrical conductivity graph)%%%%%%%%%%%%%%%%%%%%%%%%%%%%%%%%
%
    \begin{figure*}
    \centering
    \includegraphics[scale= 0.3]{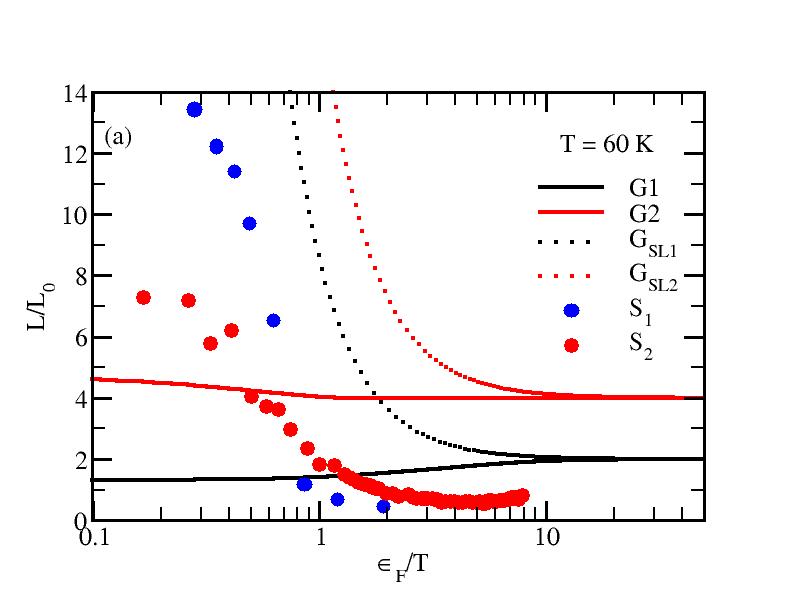}
    \includegraphics[scale= 0.3]{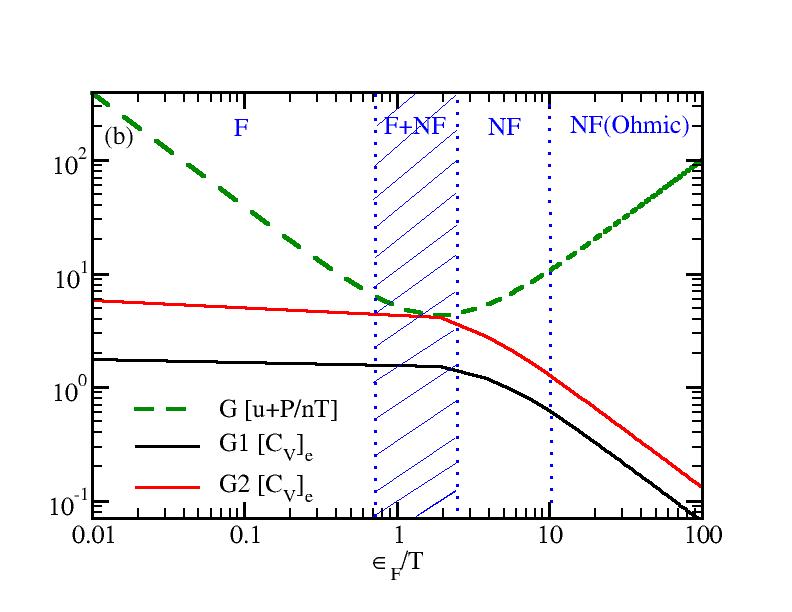}
    \caption{Lorenz ratio of 2D-G system by using fluid-type expression versus $\ep_F/T$ (LEFT) and comparison of specific heat and fluid aspect in 2D-G system versus $\ep_F/T$ (RIGHT)}
    \label{fig:my_label3}
    \end{figure*}

\begin{figure*}
   \centering
    \includegraphics[scale= 0.3]{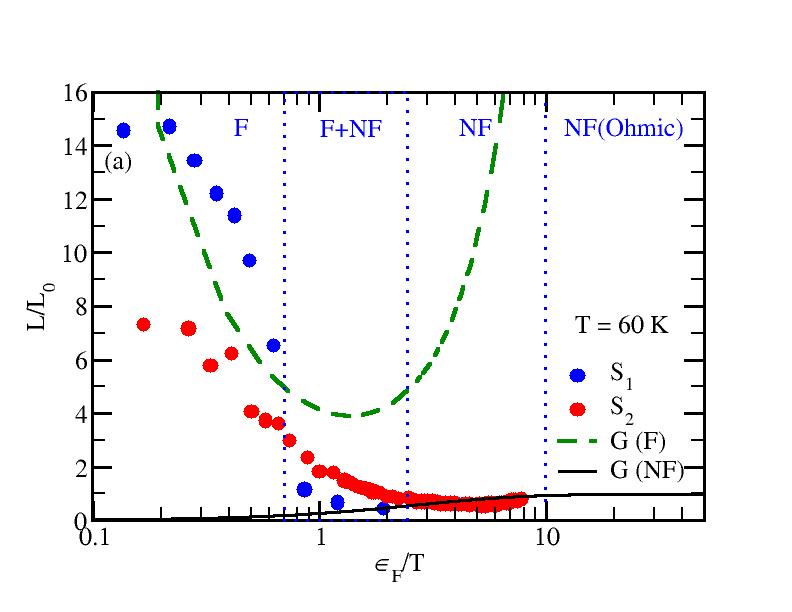}
    \includegraphics[scale= 0.3]{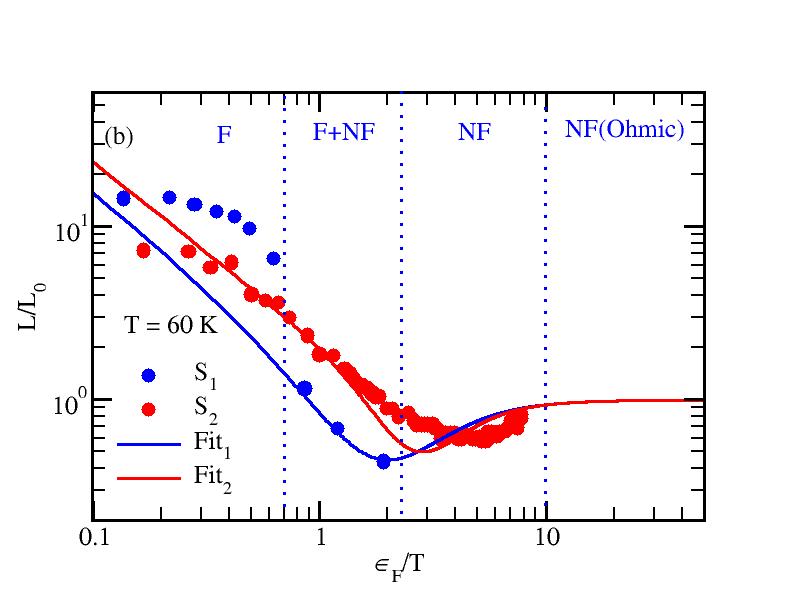}
    \caption{Lorenz ratio of fluid and non-fluid for 2D-G system versus $\ep_F/T$ (LEFT) and Lorenz ratio of the mixture of fluid and non-fluid with fitting parameters for 2D-G system versus $\ep_F/T$ (RIGHT)}
\label{f_nf_fit}
\end{figure*}
    In Fig.~\ref{fig:NF_G} and Fig.~\ref{fig:NF_NR}, we explored the Lorenz ratio ($\frac{L}{L_0}$) by using non-fluid (NF) type expressions. Fig.~\ref{fig:NF_G}(a) and (b) have shown the Lorenz ratio versus $\frac{\ep_F}{T}$ for 3D-G and 2D-G systems respectively. By using the non-fluid (NF) type of Eq.~(\ref{L_3D_G}) and its Sommerfeld limit Eq.~(\ref{L_3D_G_SL}), we plotted the black solid and dotted lines in Fig.~\ref{fig:NF_G}(a). We used the specific heat  ${[C_V]_{e1}}=\frac{\partial}{\partial T}\Big(\frac{U}{N}\Big)$, which is traditionally used in solid state or non-fluid systems. We are getting $L/L_0=1$ in the limit of $\ep_F/T\gg 1$ or Fermi Liquid (FL) regime of graphene case, which is as expected for most of the metal cases. The deviation of the Lorenz ratio from 1 has been started below $\frac{\ep_F}{T}\approx10$ as we notice from the black solid line. We have also plotted the red solid line by using Eq.~(\ref{jay23}), where another definition of specific heat ${[C_V]_{e2}}$ is used in the thermal conductivity expression. In the case of SL of 3D-G (the red dotted line), the Lorenz ratio is three instead of one. For the general case (the red solid line), the Lorenz ratio deviates from this constant line $3$ below $\frac{\ep_F}{T}\approx10$. For the 2D-G system, similar kinds of black and red solid lines are plotted in Fig.~\ref{fig:NF_G}(b) by using Eq.~(\ref{jay5}) and Eq.~(\ref{jay6}). Their respective SLs by using Eq.~(\ref{jay5SL}) and Eq.~(\ref{jay6SL}), are plotted as black and red horizontal dotted lines. We considered the surface area $(S)$ in place of volume $(V)$ in the 2D system. The Lorenz ratio when we take ${[C_{V}]_{e2}}$, the ratio becomes 2 (the red-dotted line). 
    %For both general expressions (red and black solid lines), nearly beyond $\ep_F=\epsilon_F \approx 0.05$ or $\frac{\ep_F}{T} \approx 10$, get some deviations from their SL expressions.\\

    %We already known that graphene allows for manipulating carrier density by tuning the temperature and chemical potential. 

    The Fig.~\ref{fig:NF_NR}(a) and (b), similar to Fig.~\ref{fig:NF_G}, display the Lorenz ratio in terms of $\frac{\ep_F}{T}$ for non-fluid expressions of NR case. Though the real metals have a constant chemical potential within the range, $\ep_F=2-10$~eV, we imagine that it can be tunable from $0$ to $\infty$. In Fig.~\ref{fig:NF_NR}(a), we notice that the WF law is valid in the SL of the 3D metal case, expressed by Eq.~(\ref{numaanSL}) (green dot-dashed line). We have taken $T=60^\circ$ K ($\approx 0.005$ eV), so metal range $\ep_F=2$-$10$ eV will give the ratio $\frac{\ep_F}{T}=400$-$2000$ ($\gg 1$), which is quite good domain for SL. Therefore, being situated in the SL domain, metal always follows the WF law. A slight larger value $L=1.5L_0$ comes from Eq.~(\ref{vikki2SL}) (blue dotted line), when specific heat ${[C_V]_{e2}}$ is taken. 
    %The general case stated that the WF law deviated below $\frac{\ep_F}{T}\approx10$. 
    A similar kind of behavior can be seen in Fig.~\ref{fig:NF_NR}(b) for 2D-NR systems, but interestingly, SLs in both cases, given by Eqs.~(\ref{jay3SL}) and (\ref{jay4SL}) are coincided at $L=L_0$ (blue-dotted and green dot-dashed lines). It probably indicates that if one builds 2D metal systems, then it will also follow the WF law as their Fermi energy range is again expected to be situated in the SL domain. 
    %So we can say that all non-fluid NR metal systems for 2D go along with WF law validity for both descriptions of ${[C_V]_e}$ beyond $\frac{\ep_F}{T}\approx10$. According to Fig.~\ref{fig:NF_G} and Fig.~\ref{fig:NF_NR}, we can claim that the WF law holds reasonably well for all metals in constant number and volume of thermal conductivity for any dimensionality in the high doping region. \\

    Before going to the next graphical results, let us briefly analyze what we have learned from Fig.~\ref{fig:NF_G}(a), (b) and Fig.~\ref{fig:NF_NR}(a), (b). Grossly, the non-fluid type expressions of Lorenz ratios for all cases - 3D-NR, 2D-NR, 3D-G, and 2D-G are saturated near 1 in the SL range, which means that the WF law is obeyed in the SL range. Among all four cases, 3D-NR and 2D-G are realistic cases. The former is applicable to metal systems, whose Fermi energies remain constant and within the SL range, so they always follow the WF law. Latter case - the non-fluid expressions of Lorenz ratio for 2D-G can be applicable for realistic graphene system with high doping concentration ($\epsilon_F$ is quite large or $\epsilon_F/T\gg 1$), where the WF law is obeyed well again~\cite{crossno2016observation}. However, by decreasing the doping concentration, when the graphene system approaches the charge-neutral points ($\epsilon_F$ is quite small or $\epsilon_F/T\ll 1$), the WF law violation domain is observed~\cite{crossno2016observation}. According to our non-fluid expressions, the WF law violation is also expected in the $\epsilon_F/T\ll 1$ domain but the outcome $L/L_0<1$ is not matching with experimental outcome $L/L_0>1$ by Crossno {\it et al.}~\cite{crossno2016observation}. A possible explanation of $L/L_0 > 1$ in the appearance of electron hydrodynamics \cite{Theory_G_WF} property in the domain $\ep_F/T < 1$, which is popularly called the Dirac fluid domain. Present work is firstly intended to highlight the overlooked possibility of Wiedemann-Franz law violation of any systems like NR, G or UR even we consider the non-fluid expression of $\kappa$ and $\sigma$. Then, its next attempt is to show the transition from non-fluid (NF) to fluid (F) pictures in a systematic way. An approach towards the fluid description is addressed in formalism part Sec:~(\ref{sec:F}). Let us come to the numerical estimation for fluid-type formalism.

    For numerical curves of fluid-type expressions, let us focus only on the 2D-G case. Although formalism of other cases with an Appendix is provided in the present article, one may go for result generation for graphene case. Using Eqs.~(\ref{L2DG1}) and (\ref{L2DG2}), black and red solid lines are plotted in the left panel of Fig.~\ref{fig:my_label3}. They are saturated at the values of 2 and 4, respectively, in the high doping or $\epsilon_F/T\gg 1$ domain. It means that if the fluid type expression is valid in this high doping domain 2D-G system, then the WF law may not be obeyed. However, the experimental data~\cite{crossno2016observation} claims that the WF law is well obeyed in the high doping or $\epsilon_F/T\gg 1$ domain. This means that the non-fluid expression of the Lorenz ratio should be applicable to this domain instead of fluid-type expression. The SL of Eqs.~(\ref{L2DG1}) and (\ref{L2DG2}) will also be
    \begin{align}
    &\frac{L_{G1}^{2D}}{L_0} = 2 + \frac{2 \pi^2}{3} \frac{k_B^2 T^2}{\epsilon_F^2},
    \label{2DG1_SL}\\
    &\frac{L_{G2}^{2D}}{L_0} = 4 + \frac{4 \pi^2}{3} \frac{k_B^2 T^2}{\epsilon_F^2},
     \label{2DG2_SL}
    \end{align}
    which are plotted by black and red dotted lines, respectively, in the left panel of Fig.~\ref{fig:my_label3}. They blow up from their saturated values (2 and 4) when $\epsilon_F/T$ or doping concentration decreases. 
    The experimental data of Crossno \textit{et al.}~\cite{crossno2016observation} is pasted, where S1 and S2 represent the samples of the 2D-G system in terms of lower to higher values of minimum charge carrier density. Theoretically, zero charge carrier density can be reached at $\epsilon_F/T\rightarrow 0$ but experimentally, a minimum charge carrier density~\cite{crossno2016observation} will be achieved. So, from Fig.\ref{fig:my_label3}(a), we notice that our fluid-based expression of $\frac{L_{G2}^{2D}}{L_0}$, given in Eq.(\ref{L2DG1}), (\ref{L2DG2}), can't be able to explain the experimental data points. It indicates that our formalism needs next level modification. Due to this step by step development of our formalism from (standard) non-fluid to (new directional) fluid description, we put the name of Sec:~(\ref{sec:F}) as "towards the fluid description". Now if we revisit our RTA based kinetic theory formalism, then we can find that we have systematically considered the local thermalization concept of fluid. In this concept, we consider fluid as a large number fluid elements, having velocity $u^\mu$, temperature $T$ and chemical potential $\ep_F$. Now, fluid elements will be at different position and time $(x^\mu)$, so, considering $u^\mu(x^\mu)$, $T(x^\mu)$, $\ep_F(x^\mu)$ as function of $x^\mu$ can map mathematically the local thermalization concept. Since present work is only focus on thermal and electrical conductivity, so only $T(x^\mu)$ is considered and others $u^\mu(x^\mu)$ and $\ep_F(x^\mu)$ are ignored. 
    Reader can see this consideration in Eq.(\ref{eq:BTE2}), where the information of temperature gradient $\frac{\partial T}{\partial x^i}$ is entered. Now the another term specific heat $[C_V]_e$ $\approx \frac{\partial u}{\partial T}$, which is basically considered as temperature derivative of average energy of electron. This average energy of electron information can be modified into enthalpy per electron $h = \frac{u+P}{n}$ in fluid picture. Another point is that electron-hole plasma picture will be important when we go $\frac{\ep_F}{T} < 1$ domain (towards charge neutrality point). So, we should take enthalpy per net electron $h = \frac{u+P}{n}$, where $u=u_e+u_h$, $P=P_e+P_h$ and $n=n_e-n_h$ will be respectively total internal energy density, pressure and net charge density. We have adopted the replacement
    \begin{equation}
    	[C_V]_e \approx \frac{\partial u}{\partial T} \, \rightarrow  \,  \frac{h}{T} =  \frac{u+P}{nT}
\end{equation} 
and using it in Eq.~(\ref{manohar}), we will get a better fluid-type expression of thermal conductivity:
\begin{equation}
	\kappa_G^{2D} = \frac{4 \pi k_B^2\tau_c }{h^2} \left(\frac{u+P}{nT}\right)\,f_2\left(A\right) T^2
	\label{kappa_new_specific}
\end{equation} 
and Lorenz ratio:
 \begin{equation}
    \left(\frac{L_{G1}^{2D}}{L_0}\right)  = \frac{6}{\pi^2} \left[C_V\right]^{F/NF}_G \frac{f_2\left(A\right)}{f_1\left(A\right)}~.
    \label{L_L0_F}
\end{equation}

For comparison between two quantities $C_V^e$ and $\frac{h}{T}=\frac{u+P}{nT}$, we have plotted them in the right panel of Fig.\ref{fig:my_label3}(b). Next, in Fig.\ref{f_nf_fit}(a), we have drawn dashed-line by using Eq.(\ref{L_L0_F}). Here, one can notice that the Lorenz ratio shows a divergence trend in $\frac{\ep_F}{T} < 1$ domain, which in favor of experimental data $S_1$ and $S_2$. We have also plotted non-fluid(NF) expression of Lorenz ratio, given in Eq.(\ref{jay5}), which is also in favor of experimental data in $\frac{\ep_F}{T} > 1$ domain. Based on this data favoring theoretical curves, we have classified three basic regions: $(1)$ Fluid (F) domain, $(2)$ Non-Fluid (NF) domain and $(3)$ Mixed or (F+NF) domain, which are discussed one by one (below). 
\begin{itemize}
	\item Fluid (F) domain: Fluid based Lorenz ratio, given in Eq.(\ref{L_L0_F}) valid well with experimental data in the range $0.2 < \frac{\ep_F}{T} < 0.7$.
	\item Non-Fluid (NF) domain: Non-Fluid based Lorenz ratio, given in Eq.(\ref{jay5}), valid well with experimental data in the ranges (a) $2.5 < \frac{\ep_F}{T} < 10$ and (b) $10 < \frac{\ep_F}{T} < \infty$. We can called subrange (b) as ohmic non-fluid or NF(Ohmic) domain, since we get WF law or $\frac{L}{L_0}=1$ for this $\frac{\ep_F}{T} > 10$ domain, where Fermi-Dirac distribution function can be assumed as step function. However, for subrange (a), we get $\frac{L}{L_0} < 1$ because assumption of step function is not further valid here. Corresponding domains are also marked in Fig.\ref{fig:my_label3}(b), where we can notice that $C_V^e \propto \frac{T}{\ep_F}$ follows in NF(Ohmic) domain (due to validation of electron step function like distribution) but a transition from $C_V^e \propto \frac{T}{\ep_F}$ to $C_V^e = 2k_B\Bigg[3 \frac{f_3\left(A\right)}{f_2\left(A\right)} - 2 \frac{f_2\left(A\right)}{f_1\left(A\right)} \Bigg]$ is observed in NF domain \big($2.5 < \frac{\ep_F}{T} < 10$\big). Reader can understand  this transition from Eq.(\ref{1jay3}).
	\item Mixed or (F+NF) domain: The domain $0.7 < \frac{\ep_F}{T} < 2.5$ is marked as mixed or (F+NF) domain as we have to go with replacement $C_V^e \rightarrow \frac{u+P}{nT}$ or transit from NF based quantity $C_V^e$ to fluid-based quantity $\frac{u+P}{nT}$. This domain may be considered as a grey zone and  we may put an open question for further research- How this transition happens? Interestingly, reader can notice a $\frac{\ep_F}{T}$ independent trend approximately of  $C_V^e$ and $\frac{u+P}{nT}$ within this (F+NF) domain.    
\end{itemize}      
If we analyze the F domain of Fig.\ref{fig:my_label3}(b) and Fig.\ref{f_nf_fit}(a), then we can identify that the divergent trend of $\frac{u+P}{nT}$ and $\frac{L}{L_0}$ are inter-related. Net density $n$ trends to zero when $\frac{\ep_F}{T}$ trends to zero and that is why $\frac{L}{L_0} \propto \frac{1}{n}$ trends to infinite. So, we may connect the divergent trend of Lorenz ratio in Dirac fluid domain ($\frac{\ep_F}{T} < 1$) with the fluid properties of electron-hole plasma in graphene, where fluid quantity enthalpy per particle will be the main resource for that divergence. \\
To make a bridge between the F and NF domain, we have used two switching functions. 
     \begin{equation}    
        a_F \left(x\right) = \frac{1}{e^{b_F \left(x - x_F\right)} + 1}~ ,
        \label{switch_fn1}
    \end{equation}
    \begin{equation}    
         a_{NF} \left(x\right) = \frac{1}{e^{b_{NF} \left(x - x_{NF}\right)} + 1}~ .
        \label{switch_fn2}
    \end{equation}
    where $x=\frac{\ep_F}{T}$ and $b_F$, $b_{NF}$, $x_F$, $x_{NF}$ are tuning parameters. Using these switching functions, we have prescribed a sandwich expression of Lorenz ratio as
    \begin{equation}    
        \frac{L}{L_0} = \left(\frac{L}{L_0}\right)_F \times a_F +\left(\frac{L}{L_0}\right)_{NF} \times a_{NF} ~,
        \label{Correct_fact}
    \end{equation}
where NF and F Lorenz ratio can be written from Eq. (\ref{L2DG1}) and (\ref{L_L0_F}) in a compact form. 
 \begin{equation}    
	\left(\frac{L}{L_0}\right)_{F/NF}^G =  \frac{6}{\pi^2} \,  \big[C_V\big]_{F/NF} \, \frac{f_2\left(A\right)}{f_1\left(A\right)} \,  (A=e^x)~,
	\label{compact_form}
\end{equation}
\begin{equation}    
	\left(\frac{L}{L_0}\right)_{F/NF}^Q =  \frac{9}{\pi^2} \,  \big[C_V\big]_{F/NF} \, \frac{f_3\left(A\right)}{f_2\left(A\right)} ~,
	\label{compact_form}
\end{equation}

with
\begin{equation}    
	\big[C_V^e\big]_F =  \frac{u+P}{nT}~,
\end{equation}
\begin{equation}    
	\big[C_V^e\big]_{NF} =   2k_B\Bigg[3 \frac{f_3\left(A\right)}{f_2\left(A\right)} - 2 \frac{f_2\left(A\right)}{f_1\left(A\right)} \Bigg]~.
\end{equation}
Using Eq.(\ref{compact_form}), we have plotted blue and red solid curves to fit experimental data point $S_1$(blue circles) and $S_2$(red circles). For blue line, tuning parameters are taken as $x_F = 0.2$, $x_{NF} = 0.8$, $b_F = 2$, $b_{NF} = 0.7$ and for red line, tuning parameters are taken as $x_F = 0.9$, $x_{NF} = 2$, $b_F = 2$, $b_{NF} = 0.7$. By using a better version of x-dependent switching functions, one may go with better fitting of experimental data.
 \begin{figure}
	\centering
	\includegraphics[scale= 0.3]{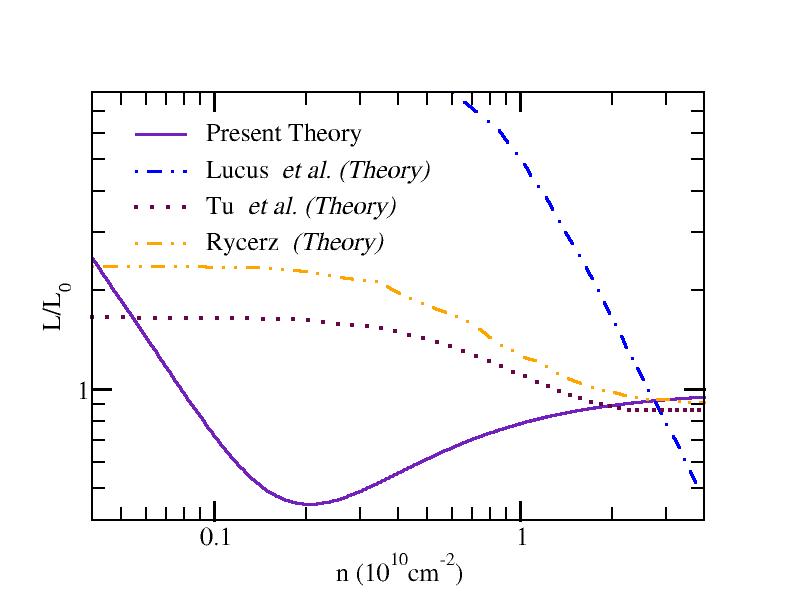}
	%\includegraphics[scale= 0.3]{L_vs_T(K).jpg}
	%\textcolor{blue}{
		\caption{Lorenz ratio of 2D-G system versus number density for different theoretical works by  Rycerz \cite{rycerz2021wiedemann}, Tu \textit{et al.} \cite{tu_Yin_23_the},  Lucus \textit{et al.} \cite{Theory_G_WF}, including present work}
		%}
	\label{fig:n_vs_LbL0Fig5}
\end{figure}
 However, instead of quantitative exact fitting, the core aim of the present work may be considered as its qualitative message or understanding on the F and NF domain and their transition. \\
This qualitative message of transition from NF to F is in agreement with earlier theoretical works by Lucas \textit{et al.}~\cite{Theory_G_WF}. Though final expressions of Lorenz ratio of Ref.~\cite{Theory_G_WF} and the present work are little different but both works have similar qualitative message- NF to F transition is associated with the WF law violation and enhancement of Lorenz ratio more than one. There are other works also which provide little different reasons for getting WF law violations. Ref.~\cite{tu_Yin_23_the} has linked this fact with presence of band gap and bipolar diffusion, while Ref.~\cite{rycerz2021wiedemann} has built thermodynamical expression of electron-hole plasma in terms of polylogarithm function, which can provide a greater than one values of Lorenz ratio. All of these theoretical results for $\frac{L}{L_0}$ along with present work are compiled in Fig.\ref{fig:n_vs_LbL0Fig5} for a comparative visibility.
%So, many more theoretical research on this topic are expected in future for getting some converging understanding. 

\begin{figure*}
		\centering
		\includegraphics[scale= 0.3]{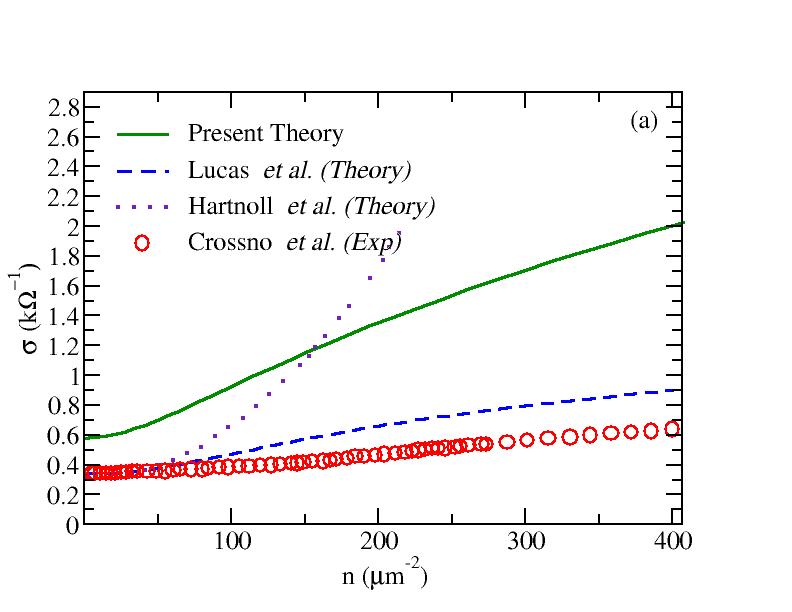}
		\includegraphics[scale= 0.3]{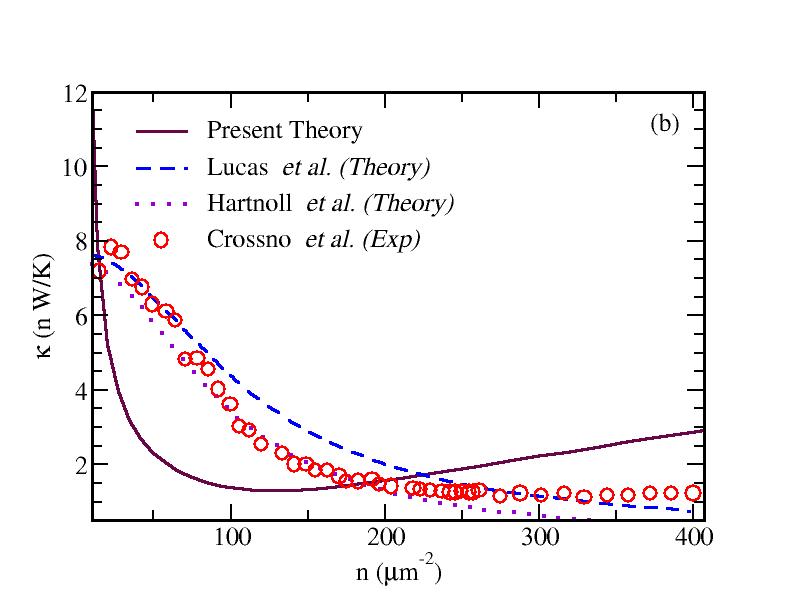}
		%\textcolor{blue}{
		\caption{Our present theoretical results of electrical conductivity (LEFT) and thermal conductivity (RIGHT) with different theoretical predictions \cite{Theory_G_WF, PhysRevB.76.144502} accompanied by experimental data ~\cite{crossno2016observation} versus number density}
		%}
		\label{Sig_K_vs_n}
	\end{figure*}

	%In this plot, we used the relaxation time approximation value of 1.1 ps, following the work of Yi Zhan.
	%
	We have done more comparison with experimental data and other earlier theory for better vissibility of present work. In Fig.\ref{Sig_K_vs_n} (a) and (b), we have plotted experimental data points (circular) of electrical conductivity $\sigma$ (in k$\Omega^{-1}$) and thermal conductivity $\kappa$ (in nW/K) from Ref.~\cite{crossno2016observation}. Here, we have added our results (solid line) by using $1$ ps relaxation time \cite{zhan2017two}, which show a non-fluid to fluid transition as we decrease the net electron density. We have also added two other theoretical results by Lucas {\it et al.} \cite{Theory_G_WF} (dash line) and Hartnoll {\it et al.} \cite{PhysRevB.76.144502} (dotted line) which are quite close to experimental data. Here, $\sigma$ of Hartnoll {\it et al.} \cite{PhysRevB.76.144502} is built via eHD framework by adding charge neutrality conductivity aspect near the charge neutrality point. Therefore, their results are quite well matched at low $n$. By extending the formalism of Ref.\cite{PhysRevB.76.144502}, Lucas {\it et al.} \cite{Theory_G_WF} additionally consider a parametrization due to local disorder and charge density and their theoretical curve of $\sigma$, $\kappa$ are quite close to experimental data. In this context, present work is focused on non-fluid to fluid transition framework, so a qualitative matching can be found. However, our future plan is to extend this NF to F transition framework into a realistic quasi-particle modeling for better fitting of experimental data of Crossno {\it et al.} \cite{crossno2016observation} as well as recent data by Majumdar {\it et al.} \cite{majumdar2025universality}. 
	%Our results also focus on the transition of NF to F with the chemical potential change and the WF law violation on the net density 
	 In future, our framework may be explored in other graphene like materials - germanene and silicene\cite{chegel2020tunable,chegel2023remarkable} as well as in dopped graphene at finite magnetic field \cite{chegel2023magneto}.
  \section{Brief Interpretation of Results}
    \label{sec:sec4}
    In this section, we will quickly revisit our results and try to interpret them briefly in terms of the transformation from non-fluid to fluid description. Here, we have gone through comparative research on Lorenz ratios for 3D, 2D NR and G systems for a systematic search of the WF law violation domain. During our analysis, we found the WF law violation domain of Fermi energy (normalized by a fixed temperature), where the Lorenz ratio deviates from one. First,
    we have used standard solid state book-based non-fluid
    expressions of Lorenz ratio for the 3D-NR case, which
    are also applied for 2D-NR, 3D-G, and 2D-G cases. For
    all cases, the Lorenz ratio remains constant in the higher
    values of normalized Fermi energy. This fact is well agreement with the universal validity of the WF law in metal systems, which can be described by the 3D-NR case of the electron gas having larger Fermi energy ($2-10$~eV) or normalized Fermi energy 400-2000 at $60^\circ$ K. Depending on the definition of specific heat at constant volume and Fermi energy, we will get two possible saturated values of Lorenz ratios - 1 and 2, where former is experimentally observed (average) value but one may mark the saturating domain as the WF law obeying domain. If one identifies two domains - less and greater than $10$ value of normalized Fermi energy (i.e. $\frac{\ep_F}{T} < 10$ and $\frac{\ep_F}{T} > 10$), then the former domain of all cases show the deviation of WF law. This tuning of normalized Fermi energy is conveniently possible for 2D-G system only via tuning the electron doping. Recently measured Lorenz ratio by Crossno \textit{et al.}~\cite{crossno2016observation} indicates the violation of the WF law in the lower normalized Fermi energy domain, which is popularly called as Dirac fluid (DF) domain. Using standard solid state or non-fluid theory for 2D-G system, we also find  WF law violation in the lower normalized Fermi energy domain. However, non-fluid theory gets less than one values of Lorenz ratio, while experiment gets its  greater than one values. It demands that we may have to transist towards a fluid-type from the standard non-fluid expressions of the Lorenz ratio. So far, to the best of our knowledge, this straightforward thermodynamical phase space analysis of Lorenz ratio, based on non-fluid theory is missing in the literature. Present work is fulfilling this gap, which may be considered one of the important contents of present article. A next level content of present article is its fluid based calculation,whose outcomes are summarized in the next paragraph.

    Towards the fluid description, we have approached in
    two steps. In the first step, we have used Boltzmann
    transport equation for 2D-G system, where local ther-
    malization concept is mainly implimented for fluid de-
    scription. If we compare the NF and F expressions of
    Lorenz ratios, then we can find that only via specific heat,
    Fermi integral functions are entered into the NF expression of Lorenz ratio. On the other hand, F expressions of
    electric and thermal conductivity and hence Lorenz ratio
    carry additional Fermi integral functions, which are com-
    pletely appeared because of local thermalization aspect
    of fluid description. During this first step attempt of fluid
    description, we get a little deviated outcomes of Lorenz
    ratio with respect to NF results. Sommerfeld limit of
    F-based Lorenz ratios give a divergent outcomes instead
    of constant values (it was found in its NF-based expression). This divergent trend of Lorenz ratio in lower nor-
    malized Fermi energy is experimentally observed, so this
    first step fluid description results may be considered as a
    preliminary hints of NF to F transition in the Dirac fluid
    domain or lower Fermi energy domain. Observing the
    quantitative mismatch between experimental data and
    first step fluid theory, we have proceeded to second step
    extension of fluid theory. Here, we have gone through the
    replacement of specific heat by enthalpy per net electron
    (divided by temperature). This replacement is based on
    the intuition for transition from quasi-particle concept
    to fluid concept. In NF picture, electron average energy
    is the quasi-particle concept and its temperature deriva-
    tive gives specific heat concept. Whereas, enthalpy per
    net electron is the fluid concept for electron-hole plasma
    in Dirac fluid domain. Interestingly, we notice that this
    conceptual transformation and replacement provide an
    enhancement of Lorenz ratio in Dirac fluid domain and
    their theoretical values are quite close to experimental
    data points.
    
  \section{Conclusion}
  \label{sec:sec5}
    Present work has explored the WF law violation of electrons in graphene in terms of a non-fluid to fluid transition framework.  After developing a step by step NF to F frameworks, we have identified four regions along the axis of normalized Fermi energy $\frac{\epsilon_F}{T}$ at $T=60^\circ$ K:
    \begin{itemize}
    	\item $\frac{\epsilon_F}{T} > 10$: Lorenz ratio is one and it can be called hardcore NF Ohmic domain. Sommerfeld limit of our NF expression can explain it easily.
    	\item Within $2.5<\frac{\epsilon_F}{T}<10$: Lorenz raio become less than one and it can be called NF but non-ohmic type domain. Our NF expression of Lorenz ratio and experimental data both hint this domain.
    	\item $0.7<\frac{\epsilon_F}{T}<2.5$: It is the transition domain from NF to F by showing less than one to greater than one values of Lorenz ratio.
    	\item $\frac{\epsilon_F}{T} < 0.7$: Here, Lorenz ratio becomes greater than one and it can be called Fluid domain as enhancing trend in fluid theory and experimental data towards lower doping range is noticed.
    \end{itemize}
    At the end, for final quantitative data matching, we have demonstrated a sandwich estimation of F and NF results. However, present article intend to draw the attention of reader on the qualitative message of NF to F transition. This association of enhancement Lorenz ratio observation in the Dirac fluid domain with the NF to F transition is agreement with earlier theoretical works. 

    \begin{acknowledgments}
    This work was partly (T.Z.W. and C.W.A.) supported by the Doctoral Fellowship in India (DIA) program of the Ministry of Education, Government of India. The authors thank the other members of eHD club Sesha P. Vempati, Ashutosh Dwibedi, Narayan Prasad, Bharat Kukkar, and Subhalaxmi Nayak.
    \end{acknowledgments}

    \appendix
    \section{{FERMI-DIRAC DISTRIBUTION FUNCTION}}
    \label{Appendix_A}
    %\begin{center}
    %    \textbf{\uppercase{A.}}
    %    \textbf{\uppercase{\underline{Fermi-Dirac distribution function}}}
    %\end{center}
    The Fermi-Dirac distribution function is
    \begin{equation}    
    f_0\left(\epsilon \right) = \frac{1}{e^{\beta \left(\epsilon - \epsilon_F\right)} + 1} = \frac{1}{A^{-1}e^{\beta \epsilon} + 1},
    \label{eq:28}
    \end{equation}
    where $\epsilon$ = energy of fermions and $\epsilon_F$ = fermi energy, and $A$ is the fugacity of the system described by
    $$ A = \exp{\left({\frac{ \epsilon_F}{k_BT}}\right)}.$$

    And the derivative of the distribution function with respect to energy is
    \begin{equation}
       -\frac{\partial f_0}{\partial \epsilon} = \frac{ \beta e^{\beta \left(\epsilon - \epsilon_F\right)}}{\left(e^{\beta \left(\epsilon - \epsilon_F\right)} + 1\right)^2} =  \frac{\partial}{\partial \epsilon_F} \left( \frac{1}{e^{\beta \left(\epsilon - \epsilon_F\right)} + 1}\right).
       \label{mintu}
    \end{equation} \\

%%%%%%%%%%%%%%%%%%

    \section{FERMI-DIRAC FUNCTION}
    \label{Appendix_B}
    %\textbf{\appendixname{-B}}

    \begin{itemize} 
    \item \textbf{The Fermi-Dirac Function}
    \begin{align}
    f_\nu (A)=\frac{1}{\Gamma (\nu)}\int_0^\infty \frac{x^{\nu-1}}{A^{-1} e^x+1} dx~,
    \label{mom}
    \end{align}
    where $f_\nu (A)$ is known as the Fermi-Dirac function.

    {\it Case.1.} When $A$ is small, then the Fermi-Dirac function can be written in a series form which is 
    \begin{align}
     f_\nu (A)= &A-\frac{A^2}{2^\nu}+\frac{A^3}{3^\nu}-\frac{A^4}{4^\nu}+ \dots \\
      & = \sum_{n=1}^{\infty} \left(-1\right)^{n-1} \frac{A^n}{n^\nu}.
    \end{align}\\

    {\it Case.2.} When $A$ is too much small $\left(\epsilon_F << k_B T\right)$, then the function becomes simplified as $A$
    \begin{equation}
    f_\nu (A) = A,
    \end{equation}
    but if we take $\epsilon_F = 0$, the function becomes unity. 

    {\it Case.3.} When the temperature is very small, and the Fermi energy has some finite value, then the Fermi-Dirac function can be written according to Sommerfeld lemma, which gives the expression of the function as
    \begin{widetext}
    \begin{equation}
    f_\nu\left(A\right) = \frac{\alpha^\nu}{\Gamma\left(\nu + 1\right)} \Bigg[1 + \nu \left(\nu -1\right)\frac{\pi^2}{6} \frac{1}{\alpha^2} + \nu \left(\nu -1\right)\left(\nu -2\right)\left(\nu -3\right)\frac{7\pi^4}{360} \frac{1}{\alpha^4}+ \dots \Bigg],
    \end{equation}
    \end{widetext}
    where $\alpha$ is given by
    $$ \alpha = \ln{A} = \frac{\epsilon_F}{k_BT}~.$$
    And
    $$ \Gamma\left(\nu + 1\right) = \nu!~, $$

    so, using this lemma, we can calculate the value of the Fermi-Dirac function for different values of $\nu$ for this particular type of case.

    {\it Case.4.} When $\left(\epsilon_F >> k_B T\right)$, then the function can be written only using the zeroth order term of Sommerfeld lemma expression.

    The derivative of Fermi function with respect to Fermi energy\\
    \begin{equation}
     \frac{\partial f_\nu \left(A\right)}{\partial \epsilon_F} = \beta  f_{\nu-1} \left(A\right).
    \end{equation}

    The derivative of Fermi function with respect to temperature\\
    \begin{equation}
    \frac{\partial f_\nu \left(A\right)}{\partial T} = \frac{1}{A}  f_{\nu-1} \left(A\right) \frac{\partial A}{\partial T}.
    \label{sd}
    \end{equation}
    The one identity for the function can be written as
    \begin{equation}
    \frac{1}{A} \frac{\partial A}{\partial T} = - \epsilon_F \beta^2 k_B = -\frac{\epsilon_F}{k_B T^2}.
    \label{rama}
    \end{equation}
    \end{itemize}

    \section{ELECTRONIC SPECIFIC HEAT}
    \label{Appendix_C}

    For 2D system graphene, which follows the linear dispersion relation, with the help of the density of state method, we can calculate the number density, total internal energy, and the electronic specific heat.

    So, the number of energy states in energy range $\epsilon$ to $\epsilon+d\epsilon$ and the surface area $S$ is written as
    \begin{equation}
    D\left(\epsilon\right) d\epsilon = g \frac{S}{\left(2\pi \hbar\right)^2} \frac{2\pi}{v_F^2} \epsilon d\epsilon.
    \end{equation}

    Now, the total number of particles at any value of temperature can be calculated as 
    \begin{equation}
    N = \int_0^{\infty}D\left(\epsilon\right)d\epsilon f_0\left(\epsilon\right).
    \label{3dnd}
    \end{equation}

    After plugging the value of $ D\left(\epsilon\right) d\epsilon $ in the above equation, we get
    \begin{equation}
    n = \frac{N}{S} = \frac{g}{2 \pi \hbar^2 v_F^2} \int_{0}^{\infty}  \frac{\epsilon}{A^{-1}e^{\beta \epsilon} + 1} d\epsilon.
    \end{equation}
    
    After solving this integration, we get the final, more general expression of number density, which is given by
    \begin{equation}
     N =  g \frac{2 \pi S }{h^2 v_F^2} \frac{\Gamma \left(2\right)}{\beta^2} f_2\left(A\right)
     \label{gauri3}
    \end{equation}
    \begin{equation}
    \implies n = \frac{k_B^2}{\pi \hbar^2 v_F^2} f_2\left(A\right) T^2~.
    \end{equation}
    
    The total internal energy of a system can be calculated as
    \begin{equation}
    U = \int_0^{\infty}  D\left(\epsilon\right)d\epsilon f_0\left(\epsilon\right) \epsilon.
    \end{equation}

    By substituting the value of $ D\left(\epsilon\right) d\epsilon $ in the above equation and solving the integration, we get the final, more general expression of total internal energy, which is given by
    \begin{equation}
    U = g \frac{ 2 \pi S }{h^2 v_F^2} \frac{\Gamma \left(3\right)}{\beta^3} f_3\left(A\right)
    \label{raman3}
    \end{equation}
    \begin{equation}
    \implies u= \frac{U}{S} = g \frac{2 \pi}{h^2 v_F^2} \frac{\Gamma \left(3\right)}{\beta^3} f_3\left(A\right).
    \label{raman4}
    \end{equation}

    Now, from the above equations, we get
    \begin{equation}
    \frac{u}{n}=\frac{U}{N} = 2 k_B T \frac{f_3\left(A\right)}{f_2\left(A\right)}.
    \label{ninad3}
    \end{equation}
    
    Let us define the specific heat capacity of the electron by taking the temperature derivative of internal energy per electron ($U/N=u/n$) while keeping surface area $S$ and $\epsilon_F$ as constants:
    \begin{align}
    \prescript{2D}{G1}{[C_{V}]_{e1}} = \Bigg[\frac{\partial }{\partial T}\Big(\frac{U}{N}\Big)\Bigg]_{S, \ \epsilon_F}.
    \end{align}
    Using the Eq. (\ref{ninad3}), the specific heat is
    \begin{align}
    \prescript{2D}{G1}{[C_{V}]_{e1}} = 2 k_B \Bigg[\frac{f_3\left(A\right)}{f_2\left(A\right)} + T \frac{\partial}{\partial T} \left( \frac{f_3\left(A\right)}{f_2\left(A\right)} \right) \Bigg].
    \label{jag3}
    \end{align}
    
    Now, the derivative part after using the identity (\ref{sd}) of the Fermi-Dirac function can be written as
    \begin{align*}
    \frac{\partial}{\partial T} \left( \frac{f_3\left(A\right)}{f_2\left(A\right)} \right) & = \frac{1}{f_2\left(A\right)} \frac{\partial}{\partial T}\left(f_3\left(A\right) \right) - f_3\left(A\right) \frac{\partial}{\partial T} \left( \frac{1}{f_2\left(A\right)} \right)\\
    & = \frac{1}{A} \frac{f_2\left(A\right)}{f_2\left(A\right)} \frac{\partial A}{\partial T} - \frac{f_3\left(A\right)}{\Big[f_2\left(A\right) \Big]^2} \frac{\partial }{\partial T} f_2\left(A\right)\\
    & = \frac{1}{A}  \frac{\partial A}{\partial T} - \frac{f_3\left(A\right)}{f_2\left(A\right)} \frac{f_1\left(A\right)}{f_2\left(A\right)} \frac{1}{A} \frac{\partial A}{\partial T}.
    \end{align*}
    The final form, we get
    %\begin{equation}
    %    \frac{1}{A}  \frac{\partial A}{\partial T} = - 2 \frac{f_2\left(A\right)}{f_1\left(A\right)} \frac{1}{T}~.
    %\end{equation}
    \begin{equation}
    \frac{\partial}{\partial T} \left( \frac{f_3\left(A\right)}{f_2\left(A\right)} \right) =  \frac{2}{T}\Bigg[\frac{f_3\left(A\right)}{f_2\left(A\right)} -  \frac{f_2\left(A\right)}{f_1\left(A\right)} \Bigg].
    \label{jeet3}
    \end{equation}

    After substituting the value from equation (\ref{jeet3}) into equation (\ref{jag3}), we get a more general form of electronic specific heat given by
    \begin{equation}
    \prescript{2D}{G1}{[C_{V}]_{e1}} = 2 k_B\Bigg[3 \frac{f_3\left(A\right)}{f_2\left(A\right)} - 2 \frac{f_2\left(A\right)}{f_1\left(A\right)} \Bigg]~.
    \label{jay3}
\end{equation} 

%%%%%%%%%%%%%%%%%%%%%%%%%%%%%%%%%
    Let us go for another definition of specific heat per electron 
    \begin{equation}
     \prescript{2D}{G2}{[C_{V}]_{e2}} = \frac{1}{N}\frac{\partial U}{\partial T}\Bigg|_{S, \ \epsilon_F}.
    \end{equation}
    From Eq.~(\ref{raman3}), the total internal energy is 
    \begin{align*}  
     &\prescript{2D}{G}{U} = \gamma T^3  f_3\left(A\right),
    \end{align*}
    where 
    $$  \gamma =  g \frac{2 \pi S }{h^2 v_F^2} \Gamma \left(3\right)k_B^3. $$
    Then, we will take the derivative of just the above equation to $T$ at constant Fermi energy $\epsilon_F$ and surface area $S$; we get
    \begin{align}
    \frac{\partial U}{\partial T}\Big|_{S,\epsilon_F} &= \gamma  \Bigg[3 T^2 f_3\left(A\right) +  T^3 f_2\left(A\right) \frac{1}{A} \frac{\partial A}{\partial T} \Bigg]\\
    & = \gamma T^2 f_2 \left(A\right) \Bigg[3 \frac{f_3\left(A\right)}{f_2\left(A\right)} +  T \frac{1}{A} \frac{\partial A}{\partial T} \Bigg]~.
      \label{bolu3}
    \end{align}
    %After using the identity (\ref{rama}), and also from the Eq.(\ref{gauri3}), the relation between $N$ and $\gamma_4$ can be written as
    %\begin{align*}
    %     \gamma_4 T^2 f_2 \left(A\right) = 2N k_B,
    %\end{align*}
    %and Eq.(\ref{bolu3}) becomes
    %\begin{align*}
    %    C_v = 2 N k_B \Bigg[3 \frac{f_3\left(A\right)}{f_2\left(A\right)} - \frac{\epsilon_F}{k_BT} \Bigg]~.
    %\end{align*}
    %Now, the final expression of electronic-specific heat per particle can be written in a more general form as
    So,
    \begin{equation}
     \prescript{2D}{G2}{[C_{V}]_{e2}}=  2k_B \Bigg[3 \frac{f_3\left(A\right)}{f_2\left(A\right)} - \frac{\epsilon_F}{k_BT} \Bigg].
    \label{sawanra3}
    \end{equation}
    
    By doing a similar type of calculation for 2D-NR case, we will get two different specific heat expressions as
    \begin{equation}
    \prescript{2D}{NR1}{[C_{V}]_{e1}} = k_B\Bigg[2 \frac{f_2\left(A\right)}{f_1\left(A\right)} -  \frac{f_1\left(A\right)}{f_0\left(A\right)} \Bigg], 
    \label{jay2}
    \end{equation}
    and
    \begin{equation}
    \prescript{2D}{NR2}{[C_{V}]_{e2}} =   k_B \Bigg[2 \frac{f_2\left(A\right)}{f_1\left(A\right)} - \frac{\epsilon_F}{k_B T}  \Bigg].
    \label{cap2}
    \end{equation}
    
    Next, for 3D-NR and 3D-G cases also, one can repeat the calculations and find the expressions of specific heat:
    \begin{equation}
    \prescript{3D}{NR1}{[C_{V}]_{e1}} = \frac{3}{2} k_B\Bigg[\frac{5}{2} \frac{f_{\frac{5}{2}}\left(A\right)}{f_{\frac{3}{2}}\left(A\right)} - \frac{3}{2} \frac{f_{\frac{3}{2}}\left(A\right)}{f_{\frac{1}{2}}\left(A\right)} \Bigg],  
    \label{jay}
    \end{equation}
    \begin{equation}
    \prescript{3D}{NR2}{[C_{V}]_{e2}} =  \frac{3}{2} k_B \Bigg[\frac{5}{2} \frac{f_{\frac{5}{2}}\left(A\right)}{f_{\frac{3}{2}}\left(A\right)} - \frac{\epsilon_F}{k_B T}  \Bigg],
    \label{dhiraj}
    \end{equation}
    \begin{equation}
    \prescript{3D}{G1}{[C_{V}]_{e1}} = 3k_B\Bigg[4 \frac{f_4\left(A\right)}{f_3\left(A\right)} - 3 \frac{f_3\left(A\right)}{f_2\left(A\right)} \Bigg], 
    \label{jay1}
    \end{equation}
    \begin{equation}
    \prescript{3D}{G2}{[C_{V}]_{e2}} =  3 k_B \Bigg[4 \frac{f_4\left(A\right)}{f_3\left(A\right)} - \frac{\epsilon_F}{k_B T}  \Bigg].
    \label{cap1}
    \end{equation}

%%%%%%%%%%%%%%%%%%%%%%%%%%%%%%%
\section{NON-FLUID DESCRIPTION of OTHER SYSTEMS}
	\label{Appendix_D}
{\bf 3D-NR:} The well-known Drude's formula for the electrical conductivity of nonrelativistic electrons in 3D solids (metals) is 
\begin{equation}
	\sigma_{NR}^{3D} = \frac{n e^2 \tau_c}{m} = \frac{n e^2 \lambda}{m v_F},
\end{equation}
and the thermal conductivity is
\begin{equation}
	\kappa_{NR}^{3D} = \frac{1}{3}n v_F \lambda \,\prescript{3D}{NR}{[C_V]_e},
\end{equation}
where, $n$ is the number (electron) density, $\tau_c$ is the relaxation time, $m$ is electron mass, $\lambda=v_F \tau_c$ is the mean free path, and $[C_V]_e$ is the electronic specific heat per particle. After taking the ratio of thermal and electrical conductivity, we get
\begin{equation}
	\frac{\kappa_{NR}^{3D}}{\sigma_{NR}^{3D}} = \frac{2}{3} \frac{\epsilon_F}{e^2} \,\prescript{3D}{NR}{[C_V]_e},
	\label{body}
\end{equation}
where $\epsilon_F = \mu = \frac{1}{2} m v_F^2$ is the Fermi energy.

Here, in Eq.~(\ref{body}), $[C_V]_e$ is the main important thing to calculate the conductivity ratio. We will use the two definitions of $[C_V]_e$: 
\begin{enumerate}
	\item $[C_{V}]_{e1}=\frac{\partial }{\partial T}\Big(\frac{U}{N}\Big)\Bigg|_{V, \ \epsilon_F}$,
	\item $[C_{V}]=\frac{\partial }{\partial T}\Big(\frac{U}{N}\Big)\Bigg|_{V, \ \epsilon_F}$,
	\item $[C_{V}]_{e2}=\frac{1}{N}\frac{\partial U}{\partial T}\Bigg|_{V, \ \epsilon_F}$.
\end{enumerate}
%
%the solid state description of specific heat $[C_v]_e=\frac{\partial U}{\partial T}\Bigg|_{N, V}$ number and volume constant $[C_{V, N}]_e$ and the fluid description of specific heat  $[C_v]_e = \frac{\partial U}{\partial T}\Bigg|_{\epsilon_F, V}$ chemical potential and volume constant $[C_{V, \epsilon_F}]_e$. 
%
%    
Here, for subsequent simplicity, we will use the notations like NR1 and G1 concerning $[C_{V}]_{e1}$ and NR2 and G2 dealing with $[C_{V}]_{e2}$. The former definition prescribes taking $T$ derivative of internal energy per particle $u=U/N$, while the latter definition says to take $T$ derivative of internal energy $U$ and then normalized by $N$. The detailed calculation can be seen in Appendix \ref{Appendix_C}. Let us first put the former specific heat,
\begin{equation}
	\prescript{3D}{NR1}{[C_{V}]_{e1}} = \frac{3}{2} k_B\Bigg[\frac{5}{2} \frac{f_{\frac{5}{2}}\left(A\right)}{f_{\frac{3}{2}}\left(A\right)} - \frac{3}{2} \frac{f_{\frac{3}{2}}\left(A\right)}{f_{\frac{1}{2}}\left(A\right)} \Bigg],
	\label{1jay}
\end{equation}
%After plugging the value of $\prescript{3D}{NR}{[C_{V, N}]_e}$ Eq.(\ref{1jay}) (details calculations can see in Appendix \ref{Appendix_C})  
in Eq.~(\ref{body}), the Lorenz ratio for 3D-NR will be
\begin{equation}
	\frac{L_{NR1}^{3D}}{L_0} = \frac{3}{\pi^2} \frac{\epsilon_F}{k_BT} \Bigg[\frac{5}{2} \frac{f_{\frac{5}{2}}\left(A\right)}{f_{\frac{3}{2}}\left(A\right)} -  \frac{3}{2} \frac{f_{\frac{3}{2}}\left(A\right)}{f_{\frac{1}{2}}\left(A\right)} \Bigg].
	\label{vikki1}
\end{equation}
%In a special limit when the temperature is very small, but Fermi energy is finite, then 
Here,
\begin{align}
	f_\nu (A)=\frac{1}{\Gamma (\nu)}\int_0^\infty \frac{x^{\nu-1}}{A^{-1} e^x+1} dx,
	\label{mom}
\end{align}
is the standard Fermi integral function (see details in Appendix \ref{Appendix_B}) with $A=e^{\epsilon_F/k_BT}$. Using Sommerfeld lemma, in the limit of $\epsilon_F/k_BT=ln A>> 1$, the electronic specific heat becomes \cite{ashcroft2022solid}
\begin{equation}
	\prescript{3D}{NR1}{[C_{V}]_{e1}} = \frac{\pi^2}{2}\frac{k_B^2 T}{\epsilon_F}, 
	\label{numaan}
\end{equation}
and then Eq.~(\ref{vikki1}) becomes
\begin{equation}
	L_{NR1}^{3D} = \frac{\pi^2}{3}\left(\frac{k_B}{e}\right)^2 = L_0,
	\label{numaanSL}
\end{equation}
which is the so-called Wiedemann-Franz law for metals.

Next, using another definition of $[C_V]_{e2}$ for the 3D-NR case, whose general expression will be (see Appendix \ref{Appendix_C})
\begin{equation}
	\prescript{3D}{NR2}{[C_{V}]_{e2}} =  \frac{3}{2} k_B \Bigg[\frac{5}{2} \frac{f_{\frac{5}{2}}\left(A\right)}{f_{\frac{3}{2}}\left(A\right)} - \frac{\epsilon_F}{k_B T}  \Bigg]~,
	\label{1dhiraj}
\end{equation}
we will get the Lorenz ratio as
\begin{equation}
	\frac{ L_{NR2}^{3D}}{L_0}  = \frac{3}{\pi^2} \frac{\epsilon_F}{k_BT} \Bigg[ \frac{5}{2} \frac{f_{\frac{5}{2}}\left(A\right)}{f_{\frac{3}{2}}\left(A\right)} - \frac{\epsilon_F}{k_B T} \Bigg].
	\label{vikki2}
\end{equation}
%In a special limit where the temperature is very small but Fermi energy is finite, 
In the limit of $\epsilon_F/T=ln A>> 1$, Sommerfeld expansion of Eq.~(\ref{1dhiraj}) becomes
\begin{equation}
	\prescript{3D}{NR2}{[C_{V}]_{e2}} = \frac{3\pi^2}{4}\frac{k_B^2 T}{\epsilon_F}, 
	\label{cv2_NR3D}
\end{equation}
and then the Eq.~(\ref{vikki2}) becomes
\begin{equation}
	L_{NR2}^{3D} =  \frac{\pi^2}{2}\left(\frac{k_B}{e}\right)^2 = 1.5 L_0.
	\label{vikki2SL}
\end{equation}
Reader can see that Eq.~(\ref{numaanSL}) and (\ref{vikki2SL}), both are independent of $T$~, $\ep_F$ and are made by universal constants like $k_B$ and $e$. So, the domain of $T-\ep_F$ plane, where $\frac{L}{L_0}$ or $L$ becomes independent of $T$~, $\ep_F$~, can be considered as Wiedemann-Franz law obeying domain and remaining will be Wiedemann-Franz law violating domain. Although, in terms of $L_0$~, $L$ is $1.5$ factor times larger for Eq.~(\ref{vikki2SL}).

{\bf 3D-G:} Now, we consider a hypothetical 3D system, which follows graphene-type dispersion relation $\epsilon=v_F p$. 
In this case, the electrical conductivity
\begin{equation}
	\sigma_{G}^{3D} = \frac{n e^2 \tau_c}{\epsilon_F}v_F^2 = \frac{n e^2 \lambda}{\epsilon_F}v_F, 
\end{equation}
and the thermal conductivity,
\begin{equation}
	\kappa_{G}^{3D} = \frac{1}{3}n v_F \lambda  \prescript{3D}{G}{[C_V]_e},
\end{equation}
will form the ratio
\begin{equation}
	\frac{\kappa_{G}^{3D}}{\sigma_{G}^{3D}} = \frac{1}{3} \frac{\epsilon_F}{e^2} \prescript{3D}{G}{[C_V]_e},
	\label{body1}
\end{equation}
where the relation between the Fermi energy and the Fermi momentum will be $\epsilon_F = p_F v_F$ for graphene.

%In Eq.(\ref{body1}), $[C_{V, N}]_e$ is coming again for 3D G case, so for number constant case,
Using two possible expressions (see Appendix \ref{Appendix_C}) of specific heat for 3D-G system,
\begin{equation}
	\prescript{3D}{G1}{[C_{V}]_{e1}} = 3k_B\Bigg[4 \frac{f_4\left(A\right)}{f_3\left(A\right)} - 3 \frac{f_3\left(A\right)}{f_2\left(A\right)} \Bigg] 
	\label{1jay1}
\end{equation}
and
\begin{equation}
	\prescript{3D}{G2}{[C_{V}]_{e2}} =  3 k_B \Bigg[4 \frac{f_4\left(A\right)}{f_3\left(A\right)} - \frac{\epsilon_F}{k_B T}  \Bigg],
	\label{1ca1}
\end{equation}
in Eq.~(\ref{body1}), we get the Lorenz ratios
\begin{equation}
	\frac{ L_{G1}^{3D}}{L_0} =  \frac{\kappa_{G1}^{3D}}{\sigma_{G1}^{3D} T L_0} = \frac{3}{\pi^2} \frac{\epsilon_F}{k_BT} \Bigg[4 \frac{f_4\left(A\right)}{f_3\left(A\right)} - 3 \frac{f_3\left(A\right)}{f_2\left(A\right)} \Bigg] 
	\label{L_3D_G}
\end{equation}
and
\begin{equation}
	\frac{ L_{G2}^{3D}}{L_0} =  \frac{\kappa_{G2}^{3D}}{\sigma_{G2}^{3D} T L_0} = \frac{3}{\pi^2} \frac{\epsilon_F}{k_BT} \Bigg[ 4\frac{ f_4\left(A\right)}{f_3\left(A\right)} - \frac{\epsilon_F}{k_B T}  \Bigg]
	\label{jay23}
\end{equation}
respectively.
In the Sommerfeld limit, Eqs.~(\ref{1jay1}) and (\ref{1ca1}) will be converted to
\begin{equation}
	\prescript{3D}{G1}{[C_{V}]_{e1}} = \pi^2\frac{k_B^2 T}{\epsilon_F},
	\label{CVe1_G}
\end{equation}
and
\begin{equation}
	\prescript{3D}{G2}{[C_{V}]_{e2}} = 3\pi^2\frac{k_B^2 T}{\epsilon_F},
	\label{CVe2_G}
\end{equation}
and so, Eqs.~(\ref{L_3D_G}) and (\ref{jay23}) become
\begin{equation}
	L_{G1}^{3D} =  \frac{\pi^2}{3}\left(\frac{k_B}{e}\right)^2 = L_0,
	\label{L_3D_G_SL}
\end{equation}
and
\begin{equation}
	L_{G2}^{3D} = \pi^2\left(\frac{k_B}{e}\right)^2 = 3 L_0,
	\label{jay23SL}
\end{equation}
respectively.

{\bf 2D-NR:} Now, let us go for 2D cases. In the case of 2D nonrelativistic system, the ratio of thermal and electrical conductivity can be written as
\begin{equation}
	\frac{\kappa_{NR}^{2D}}{\sigma_{NR}^{2D}} = \frac{\epsilon_F}{e^2} \prescript{2D}{NR}{[C_V]_e}.
	\label{body2}
\end{equation}
Using two possible expressions (see Appendix \ref{Appendix_C}) of specific heat for 2D-NR system,
\begin{equation}
	\prescript{2D}{NR1}{[C_{V}]_{e1}} = k_B\Bigg[2 \frac{f_2\left(A\right)}{f_1\left(A\right)} -  \frac{f_1\left(A\right)}{f_0\left(A\right)} \Bigg],
	\label{1jay2}
\end{equation}
and
\begin{equation}
	\prescript{2D}{NR2}{[C_{V}]_{e2}} =   k_B \Bigg[2 \frac{f_2\left(A\right)}{f_1\left(A\right)} - \frac{\epsilon_F}{k_B T}  \Bigg],
	\label{1cap2}
\end{equation}
in Eq.~(\ref{body2}), we get the Lorenz ratios
\begin{equation}
	\frac{ L_{NR1}^{2D}}{L_0} =  \frac{\kappa_{NR1}^{2D}}{\sigma_{NR1}^{2D} T L_0} = \frac{3}{\pi^2} \frac{\epsilon_F}{k_BT} \Bigg[2 \frac{f_2\left(A\right)}{f_1\left(A\right)} -  \frac{f_1\left(A\right)}{f_0\left(A\right)} \Bigg]
	\label{jay24}
\end{equation}
and
\begin{equation}
	\frac{ L_{NR2}^{2D}}{L_0} =  \frac{\kappa_{NR2}^{2D}}{\sigma_{NR2}^{2D} T L_0} = \frac{3}{\pi^2} \frac{\epsilon_F}{k_BT} \Bigg[ 2 \frac{f_2\left(A\right)}{f_1\left(A\right)} - \frac{\epsilon_F}{k_B T} \Bigg]  
	\label{jay4}
\end{equation}
respectively. In the Sommerfeld limit (SL), Eqs.~(\ref{jay24}) and (\ref{jay4}) will be converted to
\begin{equation}
	L_{NR1}^{2D} = \frac{\pi^2}{3}\left(\frac{k_B}{e}\right)^2 = L_0,
	\label{jay3SL}
\end{equation}
\begin{equation}
	L_{NR2}^{2D} = \frac{\pi^2}{3}\left(\frac{k_B}{e}\right)^2 = L_0.
	\label{jay4SL}
\end{equation}

 {\bf 3D-UR and 2D-UR:} The expressions of electrical and thermal conductivity for 3D-UR and 2D-UR will be the same as those expressions for 3D-G and 2D-G if $v_F$ is replaced by the speed of light $c$. Readers can notice that specific heat and Lorenz ratios for 3D-G and 2D-G are independent of $v_F$, so those expressions can also be used for 3D-UR and 2D-UR systems. Since our focal quantity is the Lorenz ratio in the results, we will not discuss it further. The expressions of 3D-UR and 2D-UR systems coincide with those of 3D-G and 2D-G systems. 

%%%%%%%%%%%%%%%%%%%%%%%%%%%%%%%%%
\section{APPROACH TOWARDS THE FLUID DESCRIPTION OF OTHER SYSTEMS}
\label{Appendix_E}

{\bf 2D-NR} After doing the same calculation as the 2D G of towards fluid system for a 2-dimensional nonrelativistic system, the final general expressions of the electrical and two possible thermal conductivities and their corresponding Lorenz ratios $L_{NR}^{2D}/L_0$ are given by as below:
\begin{align}
	&\sigma_{NR}^{2D} = 4\pi k_B \tau_c \left(\frac{e}{h}\right)^2 f_1\left(A\right) T,
	\label{rahul}\\
	&\kappa_{NR1}^{2D} = \frac{8\pi k_B^3 \tau_c}{h^2} \Bigg[ 2\frac{f_2\left(A\right)}{f_1\left(A\right)} - \frac{f_1\left(A\right)}{f_0\left(A\right)} \Bigg] f_2\left(A\right) T^2,
	\label{raju}\\
	& \kappa_{NR2}^{2D} = \frac{8\pi k_B^3 \tau_c}{h^2} \Bigg[2 \frac{f_2\left(A\right)}{f_1\left(A\right)} - \frac{\epsilon_F}{k_B T} \Bigg] f_2\left(A\right) T^2~,\\
	&\frac{L_{NR1}^{2D}}{L_0} = \frac{6}{\pi^2} \Bigg[ 2\frac{f_2\left(A\right)}{f_1\left(A\right)} - \frac{f_1\left(A\right)}{f_0\left(A\right)} \Bigg] \frac{f_2\left(A\right)}{f_1\left(A\right)},
	\label{Lnr2D1}\\
	& \frac{L_{NR2}^{2D}}{L_0}= \frac{6}{\pi^2} \Bigg[ 2\frac{f_2\left(A\right)}{f_1\left(A\right)} - \frac{\epsilon_F}{k_B T}  \Bigg] \frac{f_2\left(A\right)}{f_1\left(A\right)}.
	\label{vikash}
\end{align}

{\bf 3D-G:} For the 3D-G case, the expression of the electrical conductivity and two possible thermal conductivities and their corresponding Lorenz ratios $L_{G}^{3D}/L_0$ are given as below:
\begin{equation}
	\sigma_G^{3D} =  16\pi k_B^2 \tau_c \left(\frac{e^2}{3h^3 v_F}\right) f_2\left(A\right) T^2,
	\label{vinayak}
\end{equation}
\begin{equation}
	\kappa_{G1}^{3D} = \frac{48\pi k_B^4 \tau_c}{h^3 v_F} \Bigg[ 4\frac{f_4\left(A\right)}{f_3\left(A\right)} - 3 \frac{f_3\left(A\right)}{f_2\left(A\right)} \Bigg] f_3\left(A\right) T^3,
	\label{vinay}
\end{equation}
\begin{equation}
	\kappa_{G2}^{3D} = \frac{48\pi k_B^4 \tau_c}{h^3 v_F} \left[ 4\frac{f_4\left(A \right)}{f_3\left(A\right)} - \frac{\epsilon_F}{k_BT} \right] f_3\left(A\right) T^3,
	\label{vinay123}
\end{equation}
\begin{equation}
	\frac{L_{G1}^{3D}}{L_0} = \frac{27}{\pi^2} \Bigg[ 4\frac{f_4\left(A\right)}{f_3\left(A\right)} - 3 \frac{f_3\left(A\right)}{f_2\left(A\right)} \Bigg]\frac{f_3\left(A\right)}{f_2\left(A\right)},
	\label{Lg3D1}
\end{equation}
\begin{equation}
	\frac{L_{G2}^{3D}}{L_0} = \frac{27}{\pi^2} \Bigg[ 4\frac{f_4\left(A\right)}{f_3\left(A\right)} - \frac{\epsilon_F}{k_BT}\Bigg]\frac{f_3\left(A\right)}{f_2\left(A\right)}. 
	\label{Lg3D2}
\end{equation}
%
%The ratio of thermal conductivity to electrical conductivity will be 
%\begin{equation}
%   L_g^{3D} = \frac{ \left(\kappa_g^{3D}\right)_{V, N}}{\sigma_g^{3D} T} =  9 \left(\frac{k_B}{e}\right)^2 \Bigg[ 4\frac{f_4\left(A\right)}{f_3\left(A\right)} - 3 \frac{f_3\left(A\right)}{f_2\left(A\right)} \Bigg]\frac{f_3\left(A\right)}{f_2\left(A\right)}~. 
%\end{equation}
%The ratio of the Lorenz number for graphene in 3-dimension $\left(L_g^{3D}\right)$ with the Lorenz number for metal $\left(L_0\right)$ is given by the following expression
%\begin{equation}
%\left(\frac{L_g^{3D}}{L_0}\right)_{V, N} = \frac{27}{\pi^2} \Bigg[ 4\frac{f_4\left(A\right)}{f_3\left(A\right)} - 3 \frac{f_3\left(A\right)}{f_2\left(A\right)} \Bigg]\frac{f_3\left(A\right)}{f_2\left(A\right)}~. 
%  \label{Lg3D1}
%\end{equation}
%Now, at constant $V$ and $\epsilon_F$, the expression of thermal conductivity becomes
%\begin{equation}
%\left(\kappa_g^{3D}\right)_{V, \epsilon_F} = \frac{48\pi k_B^4 \tau_c}{h^3 v_F} \Bigg[ 4\frac{f_4\left(A\right)}{f_3\left(A\right)} - \frac{\epsilon_F}{k_BT}\Bigg] f_3\left(A\right) T^3~.
%\label{vinay123}
%\end{equation}
%From the equations (\ref{vinay123}) and (\ref{vinayak}), we get
%\begin{equation}
%\left(\frac{L_g^{3D}}{L_0}\right)_{V, \epsilon_F} = \frac{27}{\pi^2} \Bigg[ 4\frac{f_4\left(A\right)}{f_3\left(A\right)} - \frac{\epsilon_F}{k_BT}\Bigg]\frac{f_3\left(A\right)}{f_2\left(A\right)}~. 
%  \label{Lg3D2}
%\end{equation}

{\bf 3D-NR:} Now again, the same type of calculation is done for a 3-dimensional nonrelativistic system; we get all expressions of the electrical, thermal conductivity, and the ratio $L_{NR}^{3D}/L_0$ in a more general form, which can be expressed as
\begin{align}
	&\sigma_{NR}^{3D} = \frac{2 e^2 k_B^{\frac{3}{2}} \tau_c}{m} \left(\frac{2\pi m}{h^2}\right)^{\frac{3}{2}} f_{\frac{3}{2}}\left(A\right) T^{\frac{3}{2}}~,\\
	& \kappa_{NR1}^{3D} = \frac{15 k_B^{\frac{7}{2}} \tau_c}{2m} \left(\frac{2\pi m}{h^2}\right)^{\frac{3}{2}} \Bigg[ \frac{5}{2}\frac{f_{\frac{5}{2}}\left(A\right)}{f_{\frac{3}{2}}\left(A\right)} - \frac{3}{2}\frac{f_{\frac{3}{2}}\left(A\right)}{f_{\frac{1}{2}}\left(A\right)} \Bigg] \notag \\
	& f_{\frac{5}{2}}\left(A\right) T^{\frac{5}{2}},\\
	& \kappa_{NR2}^{3D} = \frac{15 k_B^{\frac{7}{2}} \tau_c}{2m} \left(\frac{2\pi m}{h^2}\right)^{\frac{3}{2}} \Bigg[ \frac{5}{2}\frac{f_{\frac{5}{2}}\left(A\right)}{f_{\frac{3}{2}}\left(A\right)} - \frac{\epsilon_F}{k_BT} \Bigg] \notag \\
	& f_{\frac{5}{2}}\left(A\right) T^{\frac{5}{2}},\\
	& \frac{L_{NR1}^{3D}}{L_0} = \frac{45}{4\pi^2} \Bigg[ \frac{5}{2}\frac{f_{\frac{5}{2}}\left(A\right)}{f_{\frac{3}{2}}\left(A\right)} - \frac{3}{2}\frac{f_{\frac{3}{2}}\left(A\right)}{f_{\frac{1}{2}}\left(A\right)} \Bigg] \frac{f_{\frac{5}{2}}\left(A\right)}{f_{\frac{3}{2}}\left(A\right)}~,
	\label{Lnr3D1}\\
	& \frac{L_{NR2}^{3D}}{L_0} = \frac{45}{4\pi^2} \Bigg[ \frac{5}{2}\frac{f_{\frac{5}{2}}\left(A\right)}{f_{\frac{3}{2}}\left(A\right)} - \frac{\epsilon_F}{k_BT}\Bigg] \frac{f_{\frac{5}{2}}\left(A\right)}{f_{\frac{3}{2}}\left(A\right)}.
	\label{Lnr3D2}
\end{align}
%%%%%%%%%%%%%%%%%%%%%%%%%%%%%%%%%%

%%%%%%%%%%%%%%%
%%%%%%%%%%%%%%%%%%
\bibliographystyle{unsrt}
\bibliography{ref1}

@book{ashcroft2022solid,
  title={Solid {S}tate {P}hysics},
  author={Ashcroft, Neil W and Mermin, N David},
  year={2022},
  publisher={Cengage Learning},
 url={https://www.usb.ac.ir/FileStaff/9594_2019-10-6-16-33-51.pdf}
}

@article{devanathan2021wiedemann,
  title={The {W}iedemann-{F}ranz Law for Electrical and Thermal Conduction in Metals},
  author={Devanathan, V},
  journal={J. Chennai Academy of Sciences},
  volume={4},
  pages={1--26},
  year={2021},
  url={https://www.researchgate.net/profile/V-Devanathan-2/publication/355196811_The_Wiedemann-Franz_Law_for_Electrical_and_Thermal_Conduction_in_Metals/links/616703453851f95994fc91fd/The-Wiedemann-Franz-Law-for-Electrical-and-Thermal-Conduction-in-Metals.pdf}
}

@book{Pathria:1996hda,
    author = "Pathria, R. K.",
    title = "{Statistical Mechanics}",
    edition = "2nd ed.",
    isbn = "978-0-08-054171-6",
    publisher = "Butterworth-Heinemann",
    year = "1996",
    url={https://www.pas.rochester.edu/~yishengtu/research_files/EOS_reference/Pathria,%20R.%20K.%20%20Statistical%20Mechanics.pdf}
}

@book{pillai2006solid,
  title={Solid {S}tate {P}hysics},
  author={Pillai, SO},
  year={2006},
  publisher={New Age International},
  url={https://www.google.com/search?channel=fs&client=ubuntu-sn&q=S.+Pillai%2C+Solid+state+physics+%28New+Age+International%2C+2006%29+download+pdf}
}

@book{puri1997solid,
  title={Solid {S}tate {P}hysics},
  author={Puri, RK and Babbar, VK},
  year={1997},
  publisher={S Chand and Co Ltd},
  url={https://www.academia.edu/54572618/Solid_State_Physics_Puri_Babbar/1000}
}

@article{wakeham2011gross,
  title={Gross violation of the {W}iedemann--{F}ranz law in a quasi-one-dimensional conductor},
  author={Wakeham, Nicholas and Bangura, Alimamy F and Xu, Xiaofeng and Mercure, Jean-Francois and Greenblatt, Martha and Hussey, Nigel E},
  journal={Nat. Commun.},
  volume={2},
  number={1},
  pages={396},
  year={2011},
  publisher={Nature Publishing Group UK London},
  url={https://www.nature.com/articles/ncomms1406}
}

@article{tanatar2007anisotropic,
  title={Anisotropic violation of the {W}iedemann-{F}ranz law at a quantum critical point},
  author={Tanatar, Makariy A and Paglione, Johnpierre and Petrovic, Cedomir and Taillefer, Louis},
  journal={Science},
  volume={316},
  number={5829},
  pages={1320--1322},
  year={2007},
  publisher={American Association for the Advancement of Science},
  url={https://www.jstor.org/stable/pdf/20036390.pdf}
}

@article{smith2008marginal,
  title={Marginal breakdown of the {F}ermi-{L}iquid state on the border of metallic {F}erromagnetism},
  author={Smith, RP and Sutherland, M and Lonzarich, GG and Saxena, SS and Kimura, N and Takashima, S and Nohara, M and Takagi, H},
  journal={Nature},
  volume={455},
  number={7217},
  pages={1220--1223},
  year={2008},
  publisher={Nature Publishing Group UK London},
  url={https://www.nature.com/articles/nature07401}
}

@article{pfau2012thermal,
  title={Thermal and electrical transport across a magnetic quantum critical point},
  author={Pfau, Heike and Hartmann, Stefanie and Stockert, Ulrike and Sun, Peijie and Lausberg, Stefan and Brando, Manuel and Friedemann, Sven and Krellner, Cornelius and Geibel, Christoph and Wirth, Steffen and others},
  journal={Nature},
  volume={484},
  number={7395},
  pages={493--497},
  year={2012},
  publisher={Nature Publishing Group UK London},
  url={https://www.nature.com/articles/nature11072}
}

@article{hill2001breakdown,
  title={Breakdown of {F}ermi-{L}iquid theory in a copper-oxide superconductor},
  author={Hill, RW and Proust, Cyril and Taillefer, Louis and Fournier, P and Greene, RL},
  journal={Nature},
  volume={414},
  number={6865},
  pages={711--715},
  year={2001},
  publisher={Nature Publishing Group UK London},
  url={https://www.physique.usherbrooke.ca/taillefer/Publications/2001/Article-Hill-13dec2001.pdf}
}

@article{lee2017anomalously,
  title={Anomalously low electronic thermal conductivity in metallic vanadium dioxide},
  author={Lee, Sangwook and Hippalgaonkar, Kedar and Yang, Fan and Hong, Jiawang and Ko, Changhyun and Suh, Joonki and Liu, Kai and Wang, Kevin and Urban, Jeffrey J and Zhang, Xiang and others},
  journal={Science},
  volume={355},
  number={6323},
  pages={371--374},
  year={2017},
  publisher={American Association for the Advancement of Science},
  url={https://www.jstor.org/stable/pdf/24917970.pdf}
}

@article{WP2_1_departure,
 author = "Jaoui, A. and Fauque, B. and Rischau, C. W. and others",
 title = "Departure from the {Wiedemann}-{Franz} law in ${WP_2}$ driven by mismatch in ${T}$-square resistivity prefactors",
 journal = "npj Quantum Mater. 3",
 volume = "64",
 year = {2018},
 URL={https://doi.org/10.1038/s41535-018-0136-x}
}

@article{WP2_2_departure,
  title={Thermal and electrical signatures of a hydrodynamic electron fluid in tungsten diphosphide},
  author={Gooth, Johannes and Menges, F and Kumar, N and S{\"u}$\beta$, V and Shekhar, C and Sun, Yan and Drechsler, Ute and Zierold, R and Felser, Claudia and Gotsmann, Bernd},
  journal={Nat. Commun.},
  volume={9},
  number={1},
  pages={4093},
  year={2018},
  publisher={Nature Publishing Group UK London},
  url={file:///home/iit/Downloads/s41467-018-06688-y-2.pdf}
}

@article{MoP_departure,
  title={Extremely high conductivity observed in the triple point topological metal {M}o{P}},
  author={Kumar, Nitesh and Sun, Yan and Nicklas, Michael and Watzman, Sarah J and Young, Olga and Leermakers, Inge and Hornung, Jacob and Klotz, Johannes and Gooth, Johannes and Manna, Kaustuv and others},
  journal={Nat. Commun.},
  volume={10},
  number={1},
  pages={2475},
  year={2019},
  publisher={Nature Publishing Group UK London},
  url={https://www.nature.com/articles/s41467-019-10126-y}
}

@article{crossno2016observation,
  title={Observation of the {D}irac fluid and the breakdown of the {W}iedemann-{F}ranz law in graphene},
  author={Crossno, Jesse and Shi, Jing K and Wang, Ke and Liu, Xiaomeng and Harzheim, Achim and Lucas, Andrew and Sachdev, Subir and Kim, Philip and Taniguchi, Takashi and Watanabe, Kenji and others},
  journal={Science},
  volume={351},
  number={6277},
  pages={1058--1061},
  year={2016},
  publisher={American Association for the Advancement of Science},
  url={https://www.science.org/doi/full/10.1126/science.aad0343}
}

@article{Landau1,
    author = "L D. Landau.",
    title = "{The Theory of a Fermi Liquid}",
    journal = "J. Exptl. Theoret. Phys.",
    volume = "3",
    number = "6",
    pages = "920--925",
    year = "1957",
    url={http://www.jetp.ras.ru/cgi-bin/dn/e_003_06_0920.pdf}
}

@article{Landau2,
    author = "L D. Landau.",
    title = "{Oscillations in a Fermi Liquid}",
    journal = "J. Exptl. Theoret. Phys.",
    volume = "5",
    number = "1",
    pages = "101--108",
    year = "1957",
    url={http://www.jetp.ras.ru/cgi-bin/dn/e_005_01_0101.pdf}
}

@article{Landau3,
    author = "L D. Landau.",
    title = "{On the theory of Fermi Liquid}",
    journal = "J. Exptl. Theoret. Phys. ",
    volume = "35 (8)",
    number = "6",
    pages = "70--74",
    year = "1959",
    url={http://jetp.ras.ru/cgi-bin/dn/e_008_01_0070.pdf}
}

@article{rycerz2021wiedemann,
  title={{W}iedemann--{F}ranz law for massless {D}irac fermions with implications for graphene},
  author={Rycerz, Adam},
  journal={Materials},
  volume={14},
  number={11},
  pages={2704},
  year={2021},
  publisher={MDPI},
  url={https://www.mdpi.com/1996-1944/14/11/2704}
}

@article{tu_Yin_23_the,
  title={{W}iedemann-{F}ranz law in graphene},
  author={Tu, Yi-Ting and Sarma, Sankar Das},
  journal={Phys. Rev. B},
  volume={107},
  number={8},
  pages={085401},
  year={2023},
  publisher={APS},
  url={https://journals.aps.org/prb/pdf/10.1103/PhysRevB.107.085401}
}

@article{Theory_G_WF,
  title={Transport in inhomogeneous quantum critical fluids and in the {D}irac fluid in graphene},
  author={Lucas, Andrew and Crossno, Jesse and Fong, Kin Chung and Kim, Philip and Sachdev, Subir},
  journal={Phys. Rev. B},
  volume={93},
  number={7},
  pages={075426},
  year={2016},
  publisher={APS},
  url={https://journals.aps.org/prb/pdf/10.1103/PhysRevB.93.075426}
}

@article{zhan2017two,
	title={Two distinct methods to evaluate graphene relaxation time and mobility in {B}oltzmann diffusive transport, considering ionized impurity scattering and {T}homas-{F}ermi screening},
	author={Zhan, Yi},
	journal={arXiv preprint arXiv:1712.08965},
	year={2017}
}

@article{Mahajan:2013cja,
    author = "Mahajan, Raghu and Barkeshli, Maissam and Hartnoll, Sean A.",
    title = "{Non-Fermi liquids and the {W}iedemann-{F}ranz law}",
    primaryClass = "cond-mat.str-el",
    doi = "10.1103/PhysRevB.88.125107",
    journal = "Phys. Rev. B",
    volume = "88",
    pages = "125107",
    year = "2013"
}

@article{Lucas:2018kwo,
    author = "Lucas, Andrew and Das Sarma, Sankar",
    title = "{Electronic hydrodynamics and the breakdown of the {W}iedemann-{F}ranz and Mott laws in interacting metals}",
   primaryClass = "cond-mat.str-el",
    doi = "10.1103/PhysRevB.97.245128",
    journal = "Phys. Rev. B",
    volume = "97",
    number = "24",
    pages = "245128",
    year = "2018"
}

@article{exp1,
    author = "Ku, Mark J. H. and others",
    title = "{Imaging viscous flow of the Dirac {F}luid in graphene}",
     primaryClass = "cond-mat.mes-hall",
    doi = "10.1038/s41586-020-2507-2",
    journal = "Nature",
    volume = "583",
    number = "7817",
    pages = "537--541",
    year = "2020"
}

@article{exp2,
  title={Electron hydrodynamics in anisotropic materials},
  author={Varnavides, Georgios and Jermyn, Adam S and Anikeeva, Polina and Felser, Claudia and Narang, Prineha},
  journal={Nat. Commun.},
  volume={11},
  number={1},
  pages={4710},
  year={2020},
  publisher={Nature Publishing Group UK London},
  url={https://www.nature.com/articles/s41467-020-18553-y}
}

@article{PRB_21,
  title = {Electron hydrodynamics of two-dimensional anomalous Hall materials},
  author = {Hasdeo, Eddwi H. and Ekstr\"om, Johan and Idrisov, Edvin G. and Schmidt, Thomas L.},
  journal = {Phys. Rev. B},
  volume = {103},
  issue = {12},
  pages = {125106},
  numpages = {10},
  year = {2021},
  month = {Mar},
  publisher = {American Physical Society},
  doi = {10.1103/PhysRevB.103.125106},
  url = {https://link.aps.org/doi/10.1103/PhysRevB.103.125106}
}

@article{PRB_2_21,
  title = {Beyond {O}hm's law: Bernoulli effect and streaming in electron hydrodynamics},
  author = {Hui, Aaron and Oganesyan, Vadim and Kim, Eun-Ah},
  journal = {Phys. Rev. B},
  volume = {103},
  issue = {23},
  pages = {235152},
  numpages = {13},
  year = {2021},
  month = {Jun},
  publisher = {American Physical Society},
  doi = {10.1103/PhysRevB.103.235152},
  url = {https://link.aps.org/doi/10.1103/PhysRevB.103.235152}
}

@article{PRB_3_21,
  title = {Electron-phonon hydrodynamics},
  author = {Huang, Xiaoyang and Lucas, Andrew},
  journal = {Phys. Rev. B},
  volume = {103},
  issue = {15},
  pages = {155128},
  numpages = {26},
  year = {2021},
  month = {Apr},
  publisher = {American Physical Society},
  doi = {10.1103/PhysRevB.103.155128},
  url = {https://link.aps.org/doi/10.1103/PhysRevB.103.155128}
}

@article{PRR,
  title = {Artificial electric field and electron hydrodynamics},
  author = {Tavakol, Omid and Kim, Yong Baek},
  journal = {Phys. Rev. Res.},
  volume = {3},
  issue = {1},
  pages = {013290},
  numpages = {14},
  year = {2021},
  month = {Mar},
  publisher = {American Physical Society},
  doi = {10.1103/PhysRevResearch.3.013290},
  url = {https://link.aps.org/doi/10.1103/PhysRevResearch.3.013290}
}

@article{Nat_Com_21,
    author = "Di Sante, Domenico and Erdmenger, Johanna and Greiter, Martin and Matthaiakakis, Ioannis and Meyer, Ren\'e and Rodr\'\i{}guez Fern\'andez, David and Thomale, Ronny and van Loon, Erik and Wehling, Tim",
    title = "{Turbulent hydrodynamics in strongly correlated Kagome metals}",
   primaryClass = "cond-mat.str-el",
    doi = "10.1038/s41467-020-17663-x",
    journal = "Nat. Commun.",
    volume = "11",
    number = "1",
    pages = "3997",
    year = "2020"
}

@article{PRB_coll,
  title = {Hydrodynamic collective modes in graphene},
  author = {Narozhny, B. N. and Gornyi, I. V. and Titov, M.},
  journal = {Phys. Rev. B},
  volume = {103},
  issue = {11},
  pages = {115402},
  numpages = {16},
  year = {2021},
  month = {Mar},
  publisher = {American Physical Society},
  doi = {10.1103/PhysRevB.103.115402},
  url = {https://link.aps.org/doi/10.1103/PhysRevB.103.115402}
}

@article{Nat_19,
  title={Visualizing {P}oiseuille flow of hydrodynamic electrons},
  author={Sulpizio, Joseph A and Ella, Lior and Rozen, Asaf and Birkbeck, John and Perello, David J and Dutta, Debarghya and Ben-Shalom, Moshe and Taniguchi, Takashi and Watanabe, Kenji and Holder, Tobias and others},
  journal={Nature},
  volume={576},
  number={7785},
  pages={75--79},
  year={2019},
  publisher={Nature Publishing Group UK London},
  url={https://www.nature.com/articles/s41586-019-1788-9}
}

@article{Sci_19,
  title={Quantum-critical conductivity of the {D}irac {F}luid in graphene},
  author={Gallagher, Patrick and Yang, Chan-Shan and Lyu, Tairu and Tian, Fanglin and Kou, Rai and Zhang, Hai and Watanabe, Kenji and Taniguchi, Takashi and Wang, Feng},
  journal={Science},
  volume={364},
  number={6436},
  pages={158--162},
  year={2019},
  publisher={American Association for the Advancement of Science},
  url={https://sci-hub.se/10.1126/science.aat8687}
}

@article{Hall_Sci_19,
author = {A. I. Berdyugin  and S. G. Xu  and F. M. D. Pellegrino  and R. Krishna Kumar  and A. Principi  and I. Torre  and M. Ben Shalom  and T. Taniguchi  and K. Watanabe  and I. V. Grigorieva  and M. Polini  and A. K. Geim  and D. A. Bandurin },
title = {Measuring Hall viscosity of graphene’s electron fluid},
journal = {Science},
volume = {364},
number = {6436},
pages = {162-165},
year = {2019},
doi = {10.1126/science.aau0685}
}

@article{Nat_Tec_19,
  title={Simultaneous voltage and current density imaging of flowing electrons in two dimensions},
  author={Ella, Lior and Rozen, Asaf and Birkbeck, John and Ben-Shalom, Moshe and Perello, David and Zultak, Johanna and Taniguchi, Takashi and Watanabe, Kenji and Geim, Andre K and Ilani, Shahal and others},
  journal={Nat. Nanotechnol.},
  volume={14},
  number={5},
  pages={480--487},
  year={2019},
  publisher={Nature Publishing Group UK London},
  url={https://arxiv.org/pdf/1810.10744.pdf}
}

@article{Nat_Com_18,
  title={Fluidity onset in graphene},
  author={Bandurin, Denis A and Shytov, Andrey V and Levitov, Leonid S and Kumar, Roshan Krishna and Berdyugin, Alexey I and Ben Shalom, Moshe and Grigorieva, Irina V and Geim, Andre K and Falkovich, Gregory},
  journal={Nat. Commun.},
  volume={9},
  number={1},
  pages={4533},
  year={2018},
  publisher={Nature Publishing Group UK London},
  url={https://www.nature.com/articles/s41467-018-07004-4}
}

@article{Sci_16_Res,
author = {D. A. Bandurin  and I. Torre  and R. Krishna Kumar  and M. Ben Shalom  and A. Tomadin  and A. Principi  and G. H. Auton  and E. Khestanova  and K. S. Novoselov  and I. V. Grigorieva  and L. A. Ponomarenko  and A. K. Geim  and M. Polini },
title = {Negative local resistance caused by viscous electron backflow in graphene},
journal = {Science},
volume = {351},
number = {6277},
pages = {1055-1058},
year = {2016},
doi = {10.1126/science.aad0201},
}

@ARTICLE{TZW,
	author    = "Win, Thandar Zaw and Aung, Cho Win and Khandal, Gaurav and Ghosh, Sabyasachi",
	title     = "{Graphene is neither relativistic nor non-relativistic: thermodynamics aspects}",
	journal   = "Pramana",
	publisher = "Springer Science and Business Media LLC",
	volume    =  "99",
	number    =  "1",
	month     =  "Jan",
	year      =  "2025",
	copyright = "https://www.springernature.com/gp/researchers/text-and-data-mining",
	language  = "en"
}

@article{CWA,
    author = "Aung, Cho Win and Win, Thandar Zaw and Khandal, Gaurav and Ghosh, Sabyasachi",
    title = "{Shear viscosity expression for a graphene system in relaxation time approximation}",
    primaryClass = "nucl-th",
    doi = "10.1103/PhysRevB.108.235172",
    journal = "Phys. Rev. B",
    volume = "108",
    number = "23",
    pages = "235172",
    year = "2023"
}

@article{geim2007rise,
  title={The rise of graphene},
  author={Geim, Andre K and Novoselov, Konstantin S},
  journal={Nat. Mater.},
  volume={6},
  number={3},
  pages={183--191},
  year={2007},
  publisher={Nature Publishing Group UK London},
  url={https://arxiv.org/pdf/cond-mat/0702595/1000.pdf}
}

@article{vg_nature,
  title={Dirac cones reshaped by interaction effects in suspended graphene},
  author={Elias, DC and Gorbachev, RV and Mayorov, AS and Morozov, SV and Zhukov, AA and Blake, P and Ponomarenko, LA and Grigorieva, IV and Novoselov, KS and Guinea, F and others},
  journal={Nat. Phys.},
  volume={7},
  number={9},
  pages={701--704},
  year={2011},
  publisher={Nature Publishing Group UK London},
  url={https://www.nature.com/articles/nphys2049}
}

@book{BTE,
  title={Course of {T}heoretical {P}hysics: {P}hysical {K}inetics},
  author={Pitaevskii, LP and Lifshitz, EM and Sykes, John Bradbury},
  volume={10},
  year={2017},
  publisher={Elsevier}
}

@article{VDWHRG_2023,
	title={Conductivity, diffusivity, and violation of the {W}iedemann-{F}ranz Law in a hadron resonance gas with van - der Waals interactions},
	author={Pradhan, Kshitish Kumar and Sahu, Dushmanta and Scaria, Ronald and Sahoo, Raghunath},
	journal={Physical Review C},
	volume={107},
	number={1},
	pages={014910},
	year={2023},
	publisher={APS}
}

@article{CSPM_2019,
	title={{W}iedemann-{F}ranz law for hot QCD matter in a color string percolation scenario},
	author={Sahoo, Pragati and Sahoo, Raghunath and Tiwari, Swatantra Kumar},
	journal={Physical Review D},
	volume={100},
	number={5},
	pages={051503},
	year={2019},
	publisher={APS}
}

@article{HRG1_2019,
	title={Violation of {W}iedemann-{F}ranz Law for hot hadronic matter created at {NICA}, {FAIR} and {RHIC} energies using non-extensive statistics},
	author={Rath, Rutuparna and Tripathy, Sushanta and Chatterjee, Bhaswar and Sahoo, Raghunath and Kumar Tiwari, Swatantra and Nath, Abhishek},
	journal={The European Physical Journal A},
	volume={55},
	pages={1--10},
	year={2019},
	publisher={Springer}
}

@article{HRG2_2023,
	title={Effect of time-varying electromagnetic field on {W}iedemann-{F}ranz law in a hot hadronic matter},
	author={Singh, Kamaljeet and Dey, Jayanta and Sahoo, Raghunath and Ghosh, Sabyasachi},
	journal={Physical Review D},
	volume={108},
	number={9},
	pages={094007},
	year={2023},
	publisher={APS}
}

@article{PhysRevB.76.144502,
	title = {Theory of the {N}ernst effect near quantum phase transitions in condensed matter and in dyonic black holes},
	author = {Hartnoll, Sean A. and Kovtun, Pavel K. and M\"uller, Markus and Sachdev, Subir},
	journal = {Phys. Rev. B},
	volume = {76},
	issue = {14},
	pages = {144502},
	numpages = {23},
	year = {2007},
	month = {Oct},
	publisher = {American Physical Society},
	doi = {10.1103/PhysRevB.76.144502},
	url = {https://link.aps.org/doi/10.1103/PhysRevB.76.144502}
}

@article{PhysRevLett.115.056603,
	title = {Violation of the {W}iedemann-{F}ranz Law in Hydrodynamic Electron Liquids},
	author = {Principi, Alessandro and Vignale, Giovanni},
	journal = {Phys. Rev. Lett.},
	volume = {115},
	issue = {5},
	pages = {056603},
	numpages = {5},
	year = {2015},
	month = {Jul},
	publisher = {American Physical Society},
	doi = {10.1103/PhysRevLett.115.056603},
	url = {https://link.aps.org/doi/10.1103/PhysRevLett.115.056603}
}

@article{PhysRevB.99.085104,
	title = {{W}iedemann-{F}ranz law and {F}ermi {L}iquids},
	author = {Lavasani, Ali and Bulmash, Daniel and Das Sarma, Sankar},
	journal = {Phys. Rev. B},
	volume = {99},
	issue = {8},
	pages = {085104},
	numpages = {20},
	year = {2019},
	month = {Feb},
	publisher = {American Physical Society},
	doi = {10.1103/PhysRevB.99.085104},
	url = {https://link.aps.org/doi/10.1103/PhysRevB.99.085104}
}

@article{chegel2020tunable,
	title={Tunable electronic, optical, and thermal properties of two-dimensional germanene via an external electric field},
	author={Chegel, Raad and Behzad, Somayeh},
	journal={Scientific Reports},
	volume={10},
	number={1},
	pages={704},
	year={2020},
	publisher={Nature Publishing Group UK London}
}

@article{chegel2023remarkable,
	title={Remarkable thermopower property enhancement in two-dimensional {S}i{C} via {B} and {N} doping and magnetic field},
	author={Chegel, Raad},
	journal={Journal of Alloys and Compounds},
	volume={967},
	pages={171682},
	year={2023},
	publisher={Elsevier}
}

@article{zarenia2019breakdown,
	title={Breakdown of the {W}iedemann-{F}ranz law in {AB}-stacked bilayer graphene},
	author={Zarenia, Mohammad and Smith, Thomas Benjamin and Principi, Alessandro and Vignale, Giovanni},
	journal={Physical Review B},
	volume={99},
	number={16},
	pages={161407},
	year={2019},
	publisher={APS}
}

@article{majumdar2025universality,
	title={Universality in quantum critical flow of charge and heat in ultra-clean graphene},
	author={Majumdar, Aniket and Chadha, Nisarg and Pal, Pritam and Gugnani, Akash and Ghawri, Bhaskar and Watanabe, Kenji and Taniguchi, Takashi and Mukerjee, Subroto and Ghosh, Arindam},
	journal={arXiv preprint arXiv:2501.03193},
	year={2025}
}

@phdthesis{gochan2020fermi,
	title={{F}ermi {L}iquid Properties of Dirac Materials},
	author={Gochan, Matthew P},
	year={2020},
	school={Boston College}
}

@article{principi2015violation,
	title={Violation of the {W}iedemann-{F}ranz law in hydrodynamic electron liquids},
	author={Principi, Alessandro and Vignale, Giovanni},
	journal={Phys. Rev. Lett},
	volume={115},
	number={5},
	pages={056603},
	year={2015},
	publisher={APS}
	}

@article{chegel2023magneto,
	title={Magneto-electronic and thermopower properties of B, N and Si doped monolayer graphene},
	author={Chegel, Raad},
	journal={Diamond and Related Materials},
	volume={137},
	pages={110154},
	year={2023},
	publisher={Elsevier}
}

%\printbibliography

%
%\bibliography{}
\end{document}